\documentclass[twocolumn,rmp,floatfix]{revtex4}
\usepackage{epsfig}

\begin{document}

\title{Random--Field Ising Models of Hysteresis}
\author{James P. Sethna}
\affiliation{
Laboratory of Atomic and Solid State Physics (LASSP), Clark Hall,
Cornell University, Ithaca, NY 14853-2501, USA}
\author{Karin A. Dahmen}
\affiliation{Physics Department, University of Illinois at Urbana/Champaign,
1110 West Green Street, Urbana, IL 61801-3080, USA}
\author{Olga Perkovi\'{c}}
\affiliation{URS Corporation,
600 Montgomery Street, 26th floor, San Francisco, CA 94111-2727, USA}

% Comments: 
% For inclusion in a book "Science of Hysteresis", edited
% by Giorgio Bertotti and I. Mayergoyz. 

\begin{abstract}
This is a review article of our work on hysteresis, avalanches, and
criticality. We provide an extensive introduction
to scaling and renormalization--group ideas, and discuss analytical
and numerical results for size distributions, correlation functions,
magnetization, avalanche durations and average avalanche shapes, and
power spectra. We focus here on applications to magnetic Barkhausen noise,
and briefly discuss non-magnetic systems with hysteresis and avalanches.
\end{abstract}

\maketitle

\pagenumbering{roman}
\tableofcontents
\pagenumbering{arabic}

\newcommand{\etal}{{\em et al.}\ }

\section{Introduction}
\label{sec:Intro}

\begin{figure}[thb]
  \begin{center}
    \epsfxsize=8cm
    \epsffile{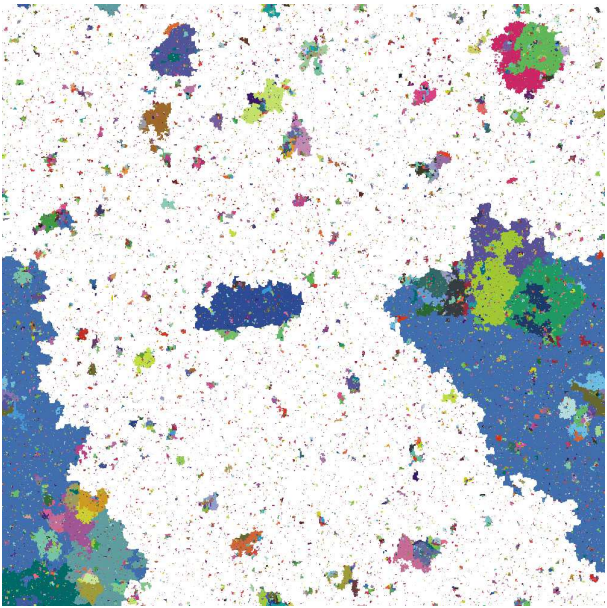}
  \end{center}
\caption{{\bf Cross section of all avalanches} in a billion--spin simulation
at the critical disorder~\cite{SethDahmMyer01}. The white background is the
infinite, spanning avalanche, which would not exist above $R_c$.
} 
\label{fig:CrossSection1000}
\end{figure}

Our group has devoted several years to the study of the dynamics of a 
simple model, the random--field Ising model at zero temperature. We do
clever simulations involving the dynamics of billions of domains 
(figure~\ref{fig:CrossSection1000}), and we develop sophisticated analytical
methods for extracting and explaining the properties of this model. 
We do this because we believe our model -- despite its dramatically
simplified nature -- may well describe the properties of Barkhausen
noise in many real physical systems. 

We present here the results of our simulations and analysis, together with
an explanation of why we believe they should (or could) describe real
experiments. Our arguments for the applicability of our model are based
on {\em renormalization group} and {\em scaling} theories, first developed
to study continuous phase transitions in equilibrium systems. To a large
extent, these theories can be seen as the underlying reason why many
if not most theories of nature apply to the real world, and (more 
specifically) why different magnets share common features in their dynamics
despite having microscopically rather different morphologies and energetics. 

\begin{figure}[thb]
  \begin{center}
    \epsfxsize=8cm
 %   \epsfbox{Fig/AvalHistoExperimentCompare.ps}
    \epsffile{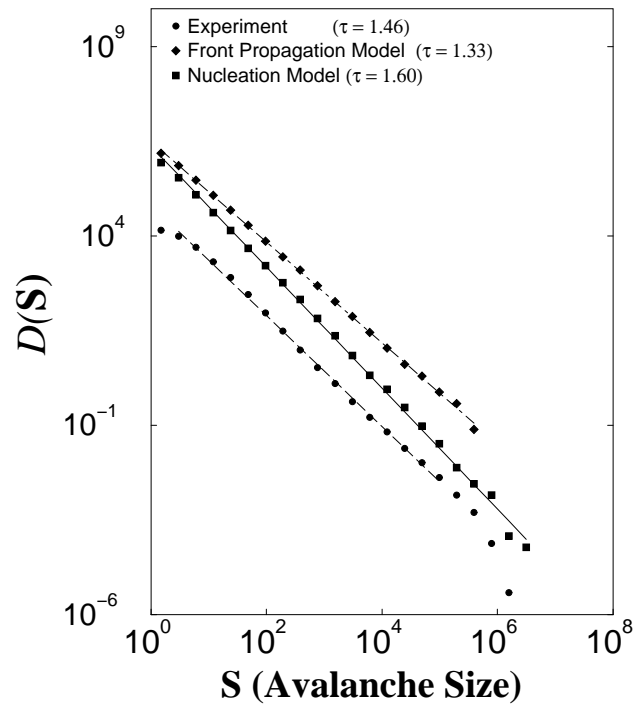}
  \end{center}
\caption{{\bf Avalanche size distributions~\cite{MehtMillDahm+02}.} The
figure shows the number $D(S)$ of avalanches of size $S$ measured in
an experiment~\cite{MehtMillDahm+02} measuring near the center of the
hysteresis loop, our (nucleation) model near $R_c$, and a related front
propagation model. Notice the straight lines on a log--log plot
indicate power--law behaviors $D(S)\sim S^{-\tau}$.
}
\label{fig:AvalHistoExperimentCompare}
\end{figure}

\begin{figure}[thb]
  \begin{center}
    \epsfxsize=8cm
    \epsffile{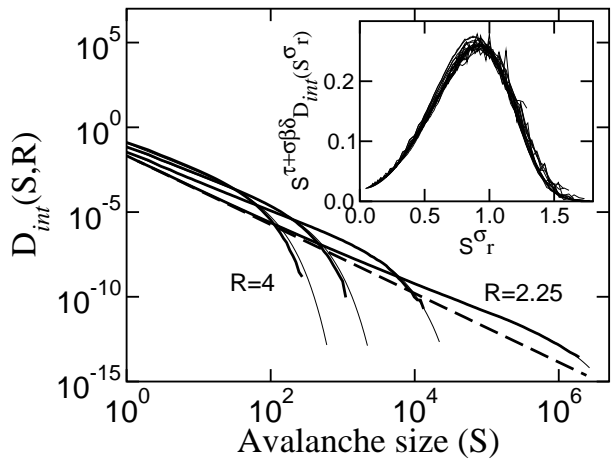}
  \end{center}
\caption{{\bf Avalanche size distributions} for our model~\cite{PerkDahmSeth95},
integrated over the hysteresis loop, for a range of disorders above $R_c=2.16$:
at $R_c$ one gets a power law (the straight dashed line) 
$D_{int}(S,R_c)\propto s^{-(\tau+\sigma\beta\delta)}$. 
The inset shows a scaling collapse of the data (equation~\ref{eq:curlyD}).
In the main figure, the thin lines show the scaling function prediction for the
associated curves. The true power--law behavior only emerges very close to 
$R_c$, but the universal scaling function governs a range of disorders 
spanning a factor of two above $R_c$. Notice also that we get six orders
of magnitude of scaling at 5\% above $R_c$, without any self--organization.
} 
\label{fig:AvalHisto}
\end{figure}

{\em Why is the noise in magnets interesting?} As one applies an external field
to most magnetic materials, the magnetization changes through the nucleation
and motion of domain walls. This motion is not smooth, and the resulting
jumps in the net magnetization is termed Barkhausen noise. These jumps,
corresponding to the reorganization (or {\em avalanche}) of a region of spins, 
usually span many decades in size: there will be many small avalanches of
spins, and fewer and fewer avalanches of larger and larger sizes (what we
call {\em crackling noise}). Indeed,
the probability $D$ of having a jump or avalanche of a given size $S$ 
often decreases as a power law%
  \footnote{In our model, this law is obtained only at the critical field
  and at the critical disorder (figure~\ref{fig:AvalHisto}: integrating 
  over field yields a different power law
  $D_{\mathrm{int}}(S) \sim S^{-(\tau + \sigma \beta \delta)}$ at the critical 
  disorder, see equation~\ref{eq:curlyD}.}
in the size of the avalanche $D(S)\sim S^{-\tau}$ for a significant range 
of sizes (figures~\ref{fig:AvalHistoExperimentCompare} and~\ref{fig:AvalHisto}).
Simple behavior (like a power law) emerging out of complicated
microscopic dynamics is
a good sign that something needs explanation! Power laws in particular
are the defining signature of continuous phase transitions, and a large
fraction of the statistical physics community focuses on problems
involving these power laws: a new class of systems to study is exciting.
Finally, similar power laws for crackling noise emerge in many other 
systems~\cite{SethDahmMyer01}, from flux--line avalanches in superconductors, 
to crumpling paper~\cite{HoulSeth96}, to earthquakes
(the Gutenberg--Richter law). Magnetic Barkhausen noise forms a manageable
experimental prototype of a whole class of behavior.

{\em Why should crackling noise in magnets be comprehensible?} Very small 
jumps in the magnetization, corresponding to avalanches on scales comparable
to the microstructure, will certainly depend upon the details
of the individual magnetic material. Large jumps, spanning the entire sample,
will depend upon the sample geometry. Indeed, these extremes in sizes are 
not regions in which simple power laws should be observed experimentally. 
But these extremes are separated by many powers of ten in a typical experiment.
Just as the complex microscopic properties of the molecules in a fluid only
affect the viscosity and density on long length scales, so also one might
expect the complicated microstructure in a magnet could be subsumed into
a few constants when dealing with events of sizes much larger than the
microstructure. The
smooth power--law behavior suggests exactly this: something simple
is happening on intermediate scales, independent of either the microscopic
or macroscopic details.

{\em What should be explicable?} How ambitious can we be in our expectations
of a successful theory? As we shall explain in this review, we cannot expect
to be able to tell when and where to expect an avalanche. 
(Our tools hence will likely not be effective at predicting large
earthquakes.) However, a successful theory should predict statistical 
averages of almost any quantity that
is dominated by events on large length and time scales, 
up to certain overall parameter--dependent scales
(analogous to viscosity and density for fluids). For example, the probability
$D(S,T,L,W,R,H)$ in our theory of having an avalanche of size $S$, 
duration $T$, long 
axis $L$, short axis $W$, at disorder $R$ and external field $H$ should be
predicted by the theory, up to an overall size scale $S_s$, time scale $T_s$, 
field scale $H_s$ and offset $H_c$, and disorder scale $R_s$ and offset $R_c$.
Indeed, we shall see that functions like these take {\em scaling forms}
\begin{eqnarray}
&&D_{\mathrm{multi}}(S,T,L,W,R,H) \cr
 &&~~~= D_s S^{-(\tau+\sigma \nu z)} 
	   \mathcal{D}_{\mathrm{multi}}(\frac{S}{S_s} r^{1/\sigma}, 
		\frac{T}{T_s} r^{z\nu}, \frac{L}{W}, \frac{h}{r^{\beta\delta}})
\end{eqnarray}
where $r = \frac{R-R_c}{R_s}$, $h = \frac{H-H_c}{H_s}$, $D_s$ is a normalization
factor, and $\tau+\sigma\nu z$
is another universal critical exponent. We shall derive these
scaling forms from self--similarity or scale invariance:
the avalanche dynamics is inherently the same when observed (say) at
microns and hundreds of microns. This scale invariance allows us to write
functions of $N$ variables as power laws times functions of $N-1$ variables;
hence, functions of one variable become pure power laws, as for $D(S)$ above.
What makes these theories predictive is that both the power law $\tau$ and
the {\em entire multivariable function} $\mathcal{D}_{\mathrm{multi}}$ are 
{\em universal}: they are independent of the microscopic model (within large
classes of systems). If your theory is in the same {\em universality class}
as the experiment, it will predict all critical exponents and scaling
functions.

There are three important qualifications about these predictions.
\begin{enumerate}
\item
{\bf Analytic corrections.} The various scales $S_s$, $T_s$, $R_c$, $H_c$,
\dots will not be constants, but will depend smoothly and analytically on 
the control parameters in a given system. This is a serious 
issue, as the function we are describing ($D_{\mathrm{multi}}$ above) will
typically itself be a smooth and analytic function of its parameters except
at a phase transition where something qualitatively changes in the behavior
of the system. Indeed, it is precisely at these phase transitions when 
events of all scales arise, where we expect our theory to apply. These
analytic corrections usually become less and less important as we confine our
attention to large, slow events near the transition.

\item
{\bf Universality Classes.} The power of the theory rests in the prediction
that quite different experimental systems, or a theoretical model and an
experiment, may be in the same universality class. We shall see that 
universality classes are studied by considering a space of possible systems:
if two systems flow towards the same fixed point as one coarse--grains to
larger scales, then their long length scale behaviors must agree. In many
cases, and for magnetic noise in particular, there can be more than one
candidate theory. Our model which incorporates short--range interactions
and nucleation of new domains has competitors which allow only interface
depinning~\cite{JiRobb92,KoilRobb00}
and which incorporate long--range fields in a mean--field fashion 
\cite{CizeZappDuri+97,UrbaMadiMark95,Nara96,ZappCizeDuri+98,DuriZapp00,
DuriZapp01}.
For many properties, the predictions of the various theories are not strikingly
different: careful experiments may be needed to distinguish between the rival
theories if it is not obvious whether nucleation or long--range interactions
are relevant in a given experiment.

\item
{\bf Dynamic criticality.} The static, equal--time properties of these models
have historically been less fussy than the time--dependent behavior. More 
specifically, there will often be several dynamic universality classes for
each static universality class, at least in the well--studied cases of
equilibrium continuous phase transitions. We will see clear evidence that
our theories are not predicting the dynamical behavior within the avalanches
properly, but that the various theories are rather successful at describing
the distributions of avalanches and other static quantities. 

\end{enumerate}

\section{Models}
\label{sec:Models}

Several variations of the zero temperature random field Ising model have
been proposed to explain the power laws in Barkhausen noise.  They are
differentiated on the basis of the presence of long range forces, and
the details of the dynamics. Our summary follows reference~\cite{KuntSeth00}.

\begin{figure}[thb]
  \begin{center}
    \epsfxsize=8cm
    \epsffile{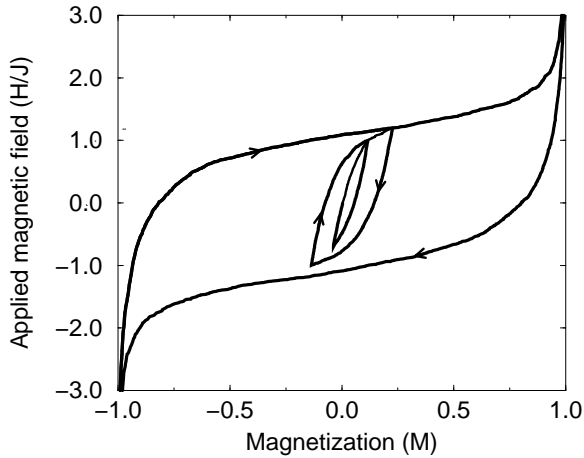}
  \end{center}
\caption{{\bf Hysteresis loop and subloops.} The magnetization in
our model (for $R>R_c$), as the external field $H$ is ramped up and down.
Our focus will primarily be on the upper, outer loop as the external field 
is ramped from $-\infty$ to $\infty$.
} 
\label{fig:HysteresisLoops}
\end{figure}

The model is composed of a large number of `spins' $s_i = \pm 1$ on a cubic
lattice (square lattice in two dimensions, hypercubic for dimensions $D>3$).
These spins represent domains or small regions of the material. The model
is subject to an external field $H$, which varies slowly in time,%
  \footnote{Except when otherwise mentioned, the
  field varies slowly enough that each avalanche finishes before the field
  changes.} 
usually ramping upward from $H=-\infty$ to $H=\infty$.
The value of a spin $s_i$ represents whether that domain is aligned ($+1$)
or anti-aligned ($-1$) with the final $H=+\infty$.
The energy function, or Hamiltonian, for the models is
\begin{eqnarray}
\label{eq:hamiltonian}
\mathcal{H} &=&-\sum_{\text{nn}} J_{\text{nn}} s_i s_j - \sum_i H s_i -
\sum_i h_i s_i \\
	&&+ \sum_i \frac{J_{\text{inf}}}{N}\,s_i - \sum_{\{i,j\}}
J_{\text{dipole}}\, \frac{3 \cos^2 (\theta_{ij}) - 1}{r_{ij}^3}\,
s_i s_j. \nonumber
\end{eqnarray}
Here $J_{\text{nn}}$ is the strength of the ferromagnetic nearest 
neighbor interactions, $h_i$ is a random field representing the effects
of compositional and morphological disorder, 
$J_{\text{inf}}$ is the strength of an infinite range demagnetizing
field,~\cite{UrbaMadiMark95} and $\theta_{ij}$ is the angle between
the positive spin direction and the difference vector between lattice 
positions $i$ and $j$, and $J_{\text{dipole}}$ is the strength of the
dipole-dipole interactions.  The critical exponents of the power laws are
independent of the
particular choice of random field distributions $\rho(h_i)$ for a large
variety of distributions. We use a Gaussian distribution of
random fields, with zero mean and standard deviation $R$.  (When we refer to
the strength of the disorder, we are referring to the width, $R$, of the
random field distribution.)%
  \footnote{In this review we focus specifically on zero temperature models;
  there are also finite temperature studies especially for pure 
  systems \cite{SideRikvNovo98,KornWhitRikvNovo01} and micromagnetic 
  systems \cite{BertBook}, interface depinning models in the presence
  of temperature \cite{RoteLubeUsad02,Midd92} and hysteresis due to driven
  interfaces at finite temperature \cite{NattPokr04,GlatNattPokr03,NattPokr+01}.
  We are currently working on hysteresis at finite temperatures with 
  randomness and nucleation~\cite{WhitDahmSeth04}.}

\begin{figure}[thb]
  \begin{center}
    \epsfxsize=8cm
    \epsffile{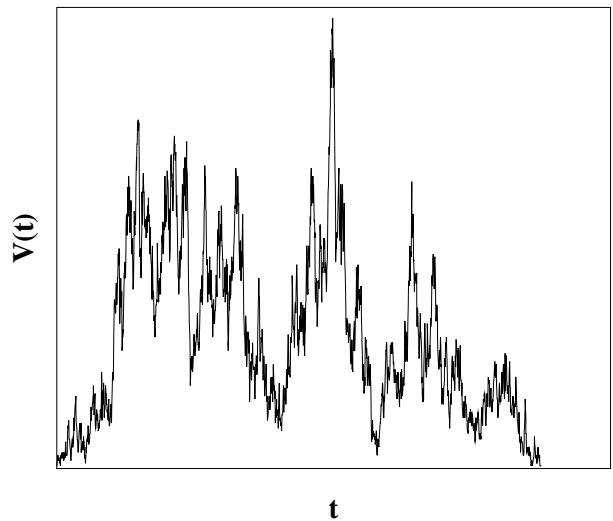}
  \end{center}
\caption{{\bf Typical avalanche time series.}~\cite{KuntSeth00}
Voltage (number of domains
flipped) pulse during a single large avalanche (arbitrary units). Notice
how the avalanche almost stops several times: if the forcing were
slightly smaller, this large avalanche would have broken up into two or
three smaller ones. The fact that the forcing is just large enough to on
average keep the avalanche growing is the cause of the self-similarity:
on average a partial avalanche of size $S$ will trigger one other on size
$S$. 
} 
\label{fig:AvalancheTypicalShape}
\end{figure}

Two different dynamics have been considered.  The first is a front
propagation dynamics in which a spin {\em on the edge of an existing front}
flips as soon as it would decrease the energy to do so, introduced by
Ji and Robbins~\cite{JiRobb92}. Spins with
no flipped neighbors cannot flip even if it would be energetically
favorable.
Second is the dynamics we use~\cite{SethDahmKart+93}, which includes
domain nucleation.
Any spin can flip when it becomes energetically favorable to do so.
In both cases, spins flip in shells---all spins which can flip at time
$t$ flip, then all of their newly flippable neighbors flip at time
$t+1$, causing an {\em avalanche}. The number of spins flipped in each
shell gives the time series
of the avalanche (figure~\ref{fig:AvalancheTypicalShape}), whose irregular
fluctuations and near halts are typical also of experimental time series.

Depending on which terms are included in the Hamiltonian, the behavior
appears to fall into three different universality classes.  

\begin{enumerate}

\item {\bf Front propagation model}~\cite{JiRobb92}.
The front propagation model has a critical field $H_c$ for a range of
disorders $R$. As the field approaches $H_c$, larger and larger avalanches
arise. As one approaches $H_c$ from below these avalanches develop
a power--law distribution in sizes with a cutoff which diverges at $H_c$.
Above $H_c$ the entire front is depinned, and the front moves forward
with an inhomogeneous velocity which begins jerky on all length and time
scales (echoing the avalanches below $H_c$) but which becomes smoother
and more uniform at large $H$. 

\begin{figure}[thb]
  \begin{center}
    \epsfxsize=8cm
    \epsffile{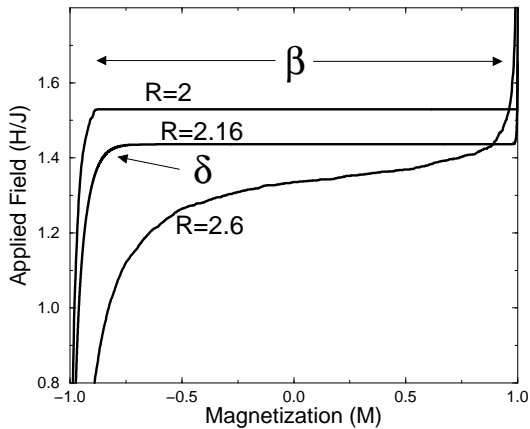}
  \end{center}
\caption{{\bf Infinite avalanche, jump in $M(H)$ for $R<R_c$.} 
} 
\label{fig:BetaDeltaJump}
\end{figure}

\item {\bf Our model (nucleation allowed, Figure~\ref{fig:HysteresisLoops}).}
For our model where domain
nucleation is allowed and only nearest neighbor interactions are
included, there is a continuous transition at a critical point
with disorder $R_c$ and external field $H_c$. 
At very large disorder, the couplings
between spins $J_{\text{nn}}$ are a small perturbation and each spin $s_i$ flips
roughly when the external field $H(t)$ is equal and opposite to the local
random field $h_i$. For all disorders above $R_c$ this qualitative behavior
persists: all avalanches are finite, dying out after a number of spin flips
that does not depend on the number of spins $N$, yielding a smooth hysteresis
loop (figure~\ref{fig:BetaDeltaJump}). For very small disorder, an early 
spin will trigger almost the entire system to flip over. This too extends
qualitatively to all disorders below $R_c$, where
finite fraction of the spins in the system (even as the system size
becomes large) flip in a single event, the {\em infinite} or {\em spanning
avalanche}. The critical disorder $R_c$ is defined as the disorder at which
the infinite avalanche first arises, and where a jump in the magnetization
per spin $m(H)$ first arises in the limit of an infinite system.%
  \footnote{Each avalanche produces a jump in the magnetization $\Delta M$,
  but all finite avalanches produce jumps in the magnetization per spin 
  $\frac{\Delta M}{N}$ which vanish as the number of spins $N$ diverges.}
Figure~\ref{fig:CrossSection1000} shows the avalanches in a cross--section
of our system at the critical disorder: the white background is the spanning,
or infinite avalanche.%
  \footnote{It may seem as if the spanning avalanche is occupying quite a large
  fraction of the volume. As the system size increases, the ``holes'' formed
  by the finite avalanches (and other spanning avalanches) will gradually
  become larger and larger fractions of the total, and the spanning avalanche
  at $R_c$ should grow with the number of spins $N$ to a power below one.}
One does not need to be exactly at $R_c$ to observe power laws in our
model: one gets a large region of power--law scaling rather far from the
critical point (figure~\ref{fig:AvalHisto} and reference~\cite{PerkDahmSeth95}).

\item {\bf Infinite Range and Dipolar Fields.}
For disorders below $R_c$, or when domain nucleation is
not allowed, the addition of an infinite-range demagnetizing
field%
  \footnote{This is not the same as a mean-field model,
   because the model contains nearest neighbor interactions along with the
   infinite range interactions.}  
\cite{UrbaMadiMark95}
can self-organize the system to a different critical behavior. 
(Self-organization means that the system
naturally sits at a critical point, without having to tune any
parameters.) The infinite-ranged interaction is sometimes introduced
to mimic the effects of the boundaries of materials with dipolar, or other
long-ranged interactions~\cite{ZappCizeDuri+98}; these interactions also 
self-organize the model to the critical point~\cite{Nara96}. Our use of the term
``infinite-range models'' perhaps
obscures the clear physical origin of this universality class of models.

Zapperi {\it et al.}~\cite{ZappCizeDuri+98} argue that the addition of
dipole-dipole interactions to the model lowers the upper
critical dimension to three and produces mean-field exponents in three
dimensions.  Since large mean-field simulations are much easier than
large simulations with dipole-dipole interactions, we will give results
from mean-field simulations in this paper. Dipolar interactions without
an infinite-range term were explored by Magni~\cite{Magn99} in two dimensions,
who found labyrinthine patterns and hysteresis loops similar to those seen in
garnet films.

The infinite--range model is apparently also equivalent to the rather
successful single--degree of freedom ABBM
models~\cite{AlesBeatBert+90e,AlesBeatBert+90t,ZappCizeDuri+98}.

\end{enumerate}

Unless we specify otherwise, we will focus on our model, allowing nucleation
with parameters $J_{\text{nn}}$, $H$, and $h_i$ but without long--range
interactions ($J_{\text{inf}} =  J_{\text{dipole}} = 0$).

\section{The Renormalization Group and Scaling}

To study crackling noise, we use renormalization-group tools
developed in the study of continuous phase transitions.%
  \footnote{This section and the next follow closely the presentation
  in reference~\cite{SethDahmMyer01}. References to the broader
  literature may be found there.} 
The word renormalization has roots in the study of quantum
electrodynamics, where the effective charge changes in size (norm) as a
function of length scale. The word group refers to the family of
coarse-graining operations basic to the method: the group product is
composition (coarsening repeatedly). The name is unfortunate, however,
as the basic coarse-graining operation does not have an inverse, and
thus the renormalization group does not have the mathematical structure
of a group.

\begin{figure}[thb]
  \begin{center}
    \epsfxsize=8cm
    \epsffile{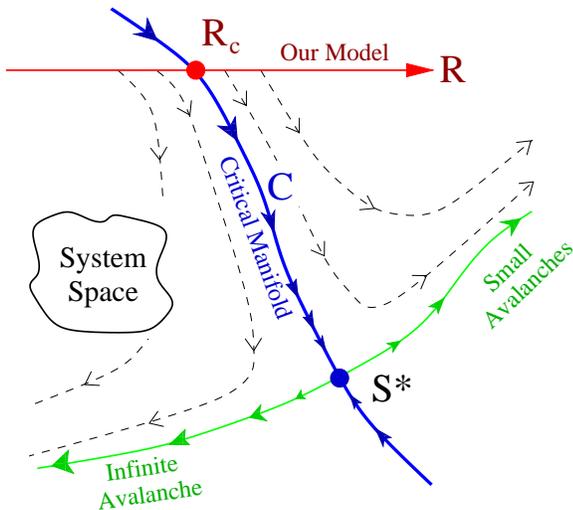}
  \end{center}
\caption{{\bf Renormalization--group flows}~\cite{SethDahmMyer01}. The
renormalization--group uses coarse--graining to longer length scales to
produce a mapping from the space of physical systems into itself. Consider
the space of all possible systems exhibiting magnetic hysteresis (including,
in our imagination, both real models and experimental systems). Each
model can be coarse--grained, removing some fraction of the
microscopic domains and introducing more complex dynamical rules so
that the remaining domains still flip over at the same external fields. 
This defines a mapping of our space of models into itself. A fixed point
$\mathbf{S^*}$ in this space will be self--similar: because it maps
into itself upon coarse--graining, it must have the same behavior on 
different length scales. Points that flow into $\mathbf{S^*}$ under
coarse--graining share this self--similar behavior on sufficiently long
length scales: they all share the same {\em universality class}.
} 
\label{fig:RGFlows}
\end{figure}

\begin{figure}[thb]
  \begin{center}
    \epsfxsize=8cm
    \epsffile{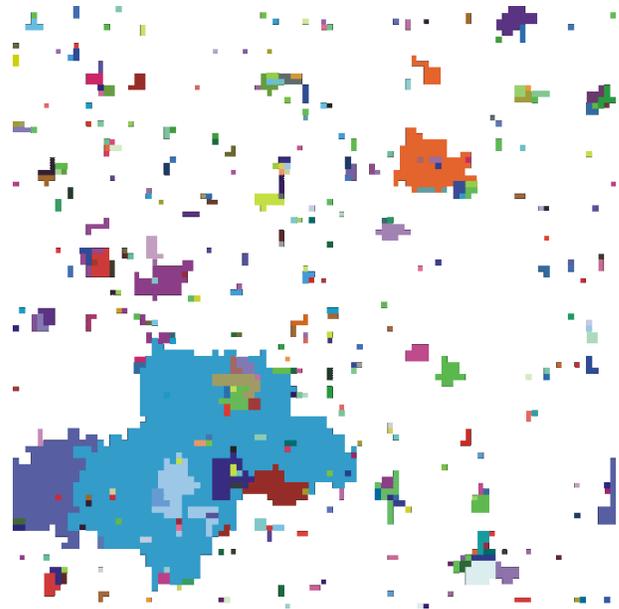}
  \end{center}
\caption{{\bf Cross section of all avalanches} in a million--spin simulation
at the critical disorder~\cite{SethDahmMyer01}. Compare with the much
larger simulation in figure~\ref{fig:CrossSection1000}.
} 
\label{fig:CrossSection100}
\end{figure}

The renormalization group studies the way the space of all physical
systems maps into itself under coarse-graining (see figure~\ref{fig:RGFlows}).
The coarse-graining operation shrinks the system, and removes degrees of
freedom on short length scales. Under coarse-graining, we often find a
fixed point $\mathbf{S^*}$: many different models flow into the fixed point and
hence share long-wavelength properties. To get a schematic view of
coarse-graining, look at figures~\ref{fig:CrossSection1000} 
and~\ref{fig:CrossSection100}: the $1000^3$ cross section looks
(statistically) like the $100^3$ section if you blur your eyes by a factor
of 10. 

Much of the mathematical complexity of this field involves
finding analytical tools for computing the flow diagram in
figure~\ref{fig:RGFlows}.
Using methods developed to study thermodynamical phase transitions and
the depinning of charge-density waves, we can calculate for our model
the flows for systems in dimensions close to six (the so-called $\epsilon$
expansion, where  $\epsilon=6-D$, $D$ being the dimension of the system).
Interpolating between dimensions may seem a surprising thing to do. In
our system it gives rather good predictions even in three dimensions
(i.e., $D=3$), but it's hard work, and we won't discuss it here. Nor will we
discuss real-space renormalization-group methods or series expansion
methods. We focus on the relatively simple task of using the
renormalization group to justify and explain the universality,
self-similarity, and scaling observed in nature.

Consider the ``system space" for disordered magnets. There is a separate
dimension in system space for each possible parameter in a theoretical
model (disorder, coupling, next-neighbor coupling, dipolar fields {\it etc.})
or in an experiment (temperature, annealing time, chemical composition
...). Coarse-graining, however one implements it, gives a mapping from
system space into itself: shrinking the system and ignoring the shortest
length scales yields a new physical system with identical long-distance
physics, but with different (renormalized) values of the parameters.
We've abstracted the problem of understanding crackling noise in magnets
into understanding a dynamical system acting on the space of all
dynamical systems.

Figure~\ref{fig:RGFlows} represents a two-dimensional cross section of
this infinite-dimensional system space. We've chosen the cross section
to include our model: as we vary the disorder $R$, our model sweeps out
a straight line in system space. The cross section also includes a
fixed point $\mathbf{S^*}$, which maps into itself under
coarse-graining.  The system $\mathbf{S^*}$ looks the same on all length
and time scales, because it coarse-grains into itself. We can picture
the cross section of figure~\ref{fig:RGFlows} either as a plane in
system space (in which case the arrows and flows depict projections,
since in general the real flows will point somewhat out of the plane),
or as the curved manifold swept out by our one-parameter model as we
coarse grain (in which case the flows above our model and below the
horizontal curved line in figure~\ref{fig:RGFlows} should be ignored).

The flow near $\mathbf{S^*}$ has one unstable direction, leading outward
along the horizontal curve (the unstable manifold). In system space, there is
a surface of points $\mathbf{C}$ which flow into $\mathbf{S^*}$ under
coarse-graining. Because $\mathbf{S^*}$ has only one unstable direction,
$\mathbf{C}$ divides system space into two phases. To the left of
$\mathbf{C}$, the systems will have one large, system-spanning avalanche
(a snapping noise). To the right of $\mathbf{C}$, all avalanches are
finite and under coarse-graining they all become small (popping noise).
Our model, as it crosses $\mathbf{C}$ at the value $R_c$, goes through a
phase transition.

Our model at $R_c$ is not self-similar on the shortest length scales (where
the square lattice of domains still is important), but because it flows
into $\mathbf{S^*}$ as we coarse-grain we deduce that it is self-similar on long
length scales. Some phase transitions, like ice melting into water, are
abrupt and don't exhibit self-similarity. Continuous phase transitions
like ours almost always have self-similar fluctuations on long length
scales. Also, we must note that our model at $R_c$ will have the same
self-similar structure as $\mathbf{S^*}$ does. Indeed, any experimental or
theoretical model lying on the critical surface $\mathbf{C}$ will share the same
long-wavelength critical behavior. This is the fundamental explanation
for universality.

\begin{figure}[thb]
  \begin{center}
    \epsfxsize=8cm
    \epsffile{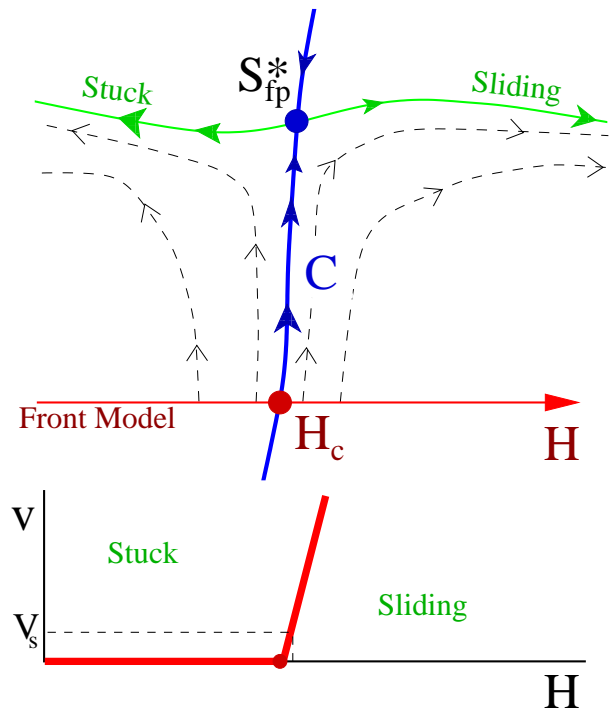}
  \end{center}
\caption{{\bf Flows for a front--propagation model}~\cite{SethDahmMyer01}.
The front propagation model has a critical field $H_c$ at which the 
front changes from a pinned to sliding state. 
(a)~Coarse-graining defines a flow on the space of depinning models.
The fixed point $\mathbf{S^*}_{fp}$
will have a different local flow field from other renormalization-group
fixed points, yielding its own universality class of critical exponents
and scaling functions. The critical manifold $\mathbf{C}$, consisting of
models which flow into $\mathbf{S^*}_{fp}$, separates the stuck fronts
from those which slide forward with an average velocity $v(H)$. 
(b)~The velocity varies with the external force as a power law 
$v(H)\sim (H-H_c)^\beta$. Clever experiments, or long--range fields, can
act to control not the external field, but the net magnetization: changing
the magnetization slowly sets $v\approx 0$, thus self--tuning $H\approx H_c$.
This is one example of {\em self-organized criticality}.
} 
\label{fig:SelfOrganized}
\end{figure}
 
\begin{figure}[thb]
  \begin{center}
    \epsfxsize=8cm
    \epsffile{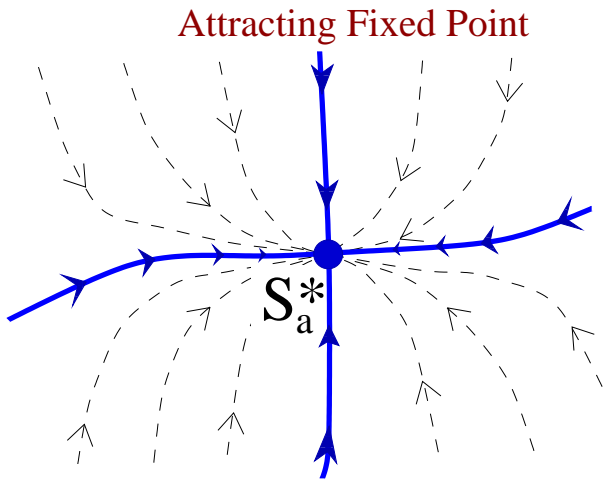}
  \end{center}
\caption{{\bf Attracting fixed point}~\cite{SethDahmMyer01}.
Often there will be fixed points that
attract in all directions. These fixed points describe phases rather
than phase transitions. Most phases are rather simple, with fluctuations
that die away on long length scales. When fluctuations remain
important, they will exhibit self-similarity and power laws called
{\em generic scale invariance}.
} 
\label{fig:GenericSI}
\end{figure}

The flows in system space can vary from one class of problems to
another: the system space for front propagation models
(figure~\ref{fig:SelfOrganized}a) will
have a different flow, and its fixed point will have different scaling
behavior (yielding a different universality class). In some cases, a
fixed point will attract all the systems in its vicinity
(figure~\ref{fig:GenericSI}).
Usually at such attracting fixed points the
fluctuations become unimportant at long length scales: the Navier-Stokes
equation for fluids described earlier can be viewed as a stable fixed
point. The coarse-graining process, averaging over many degrees of
freedom, naturally smooths out fluctuations, if they aren't amplified
near a critical point by the unstable direction. Fluctuations can remain
important when a system has random noise in a conserved property, so
that fluctuations can only die away by diffusion: in these cases, the
whole phase will have self-similar fluctuations, leading to {\em generic
scale invariance}.

Sometimes, even when the system space has an unstable direction like in
figure~\ref{fig:RGFlows}, the observed behavior always has avalanches of
all scales. This can occur simply because the physical system averages
over a range of model parameters (i.e., averaging over a range of $R$
including $R_c$ in figure~\ref{fig:RGFlows}). For example, this can
occur by the sweeping of a parameter slowly in time, or varying it
gradually in space either deliberately or through large-scale
inhomogeneities.

One can also have {\em self-organized criticality}, where the system is
controlled so that it naturally sits on the critical surface.
Self-organization to the critical point can occur via many mechanisms.
In the front propagation models with long--range fields
(figure~\ref{fig:SelfOrganized}b), upon an increment of external field 
$H$ the front will move forward until the long--range demagnetization force
balances the increment: the net external field given by the sum organizes
itself to precisely the critical depinning field for the front.

\section{Critical Exponents, Scaling Functions}   
\label{sec:ExponentsScaling}

The renormalization group is the theoretical underpinning for
understanding why universality and self-similarity occur.%
  \footnote{This section also follows closely the presentation
  in reference~\cite{SethDahmMyer01}. 
%References to the broader
%  literature may be found there.
} 
Once we grant that different systems should sometimes share long-distance
properties, though, we can quite easily derive some powerful predictions.

To take a tangible example, lets consider the relation between the
duration of an avalanche and its size.  If we
look at all avalanches of a certain duration $T$ in an experiment, they
will have a distribution of sizes $S$ around some average
$\langle S\rangle_{\mathrm{exp}}(T)$. If we look at a theoretical
model, it will have a corresponding average size 
$\langle S\rangle_{\mathrm{th}}(T)$. If our model describes the
experiment, these functions must be essentially the same at large $S$ and
large $T$. We must allow for the fact that the experimental units of time
and size will be different from the ones in our model: the best we can
hope for is that 
\begin{equation}
\langle S \rangle_{\mathrm{exp}}(T) 
	= A \langle S\rangle_{\mathrm{th}}(T/B),
\end{equation}
for some rescaling factors $A$ and $B$.

Now, instead of comparing to experiment, we can compare our model to
itself on a slightly larger time scale. If the time scale is expanded
by a small factor $B=1/(1-\delta)$, then the rescaling of the size will
also be small, say $1+a$. Hence
\begin{equation}
\langle S\rangle(T) = (1 + a\delta) \langle S\rangle((1-\delta) T).
\end{equation}
Making $\delta$ very small yields the simple relation
$a \langle S\rangle=T d\langle S\rangle/dT$, which can
be solved to give the power law relation $\langle S\rangle(T) = S_0 T^a$.
The exponent $a$
is called a critical exponent, and is a universal prediction of a given
theory. (That means that if the theory correctly describes an
experiment, the critical exponents will agree.) In our work, we write
the exponent $a$ relating time to size in terms of three other critical
exponents, $a=1/\sigma\nu z$.

\vskip 0.25truein
\begin{figure}[thb]
  \begin{center}
    \epsfxsize=8cm
    \epsffile{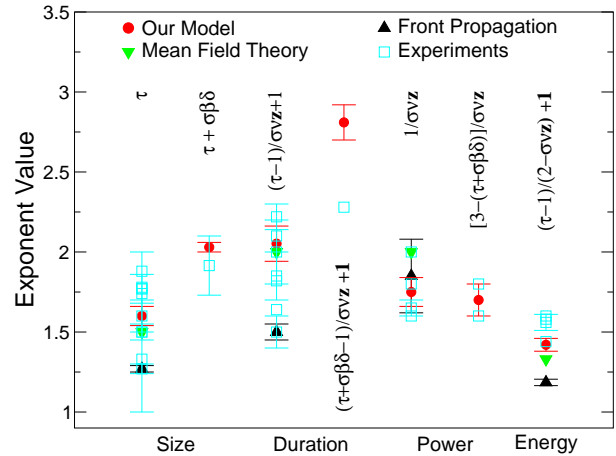}
  \end{center}
\caption{{\bf Universal Critical Exponents vs.\ 
Experiment.}~\cite{SethDahmMyer01}
Different experiments on crackling noise in magnets measure
different combinations
of the universal critical exponents. Here we compare experimental
measurements (see table I of reference~\cite{PerkDahmSeth95}) to the theoretical
predictions for three models: our model, the front-propagation
model and the dipolar mean-field theory.
Power laws giving the
probability of getting an avalanche of a given size, duration, or energy
at the critical point are shown; also shown is the critical exponent
giving the power as a function of frequency~\cite{KuntSeth00}
(due to the internal structure of the avalanches, 
figure~\ref{fig:AvalancheTypicalShape}). In each pair of columns, the
first column includes only avalanches at external fields $H$ near $H_c$ 
where the largest avalanches occur, 
and the second column (when it exists) includes all avalanches. The
various combinations of the basic critical exponents can be derived from
exponent equality calculations similar to the one discussed in the
text. Many of the experiments were done years before the
theories were developed: many did not report error bars. All three
theories do well (especially considering the possible systematic errors
in fitting power laws to the experimental measurements: see 
in figure~\ref{fig:AvalHisto} how a bit away from $R_c$ the effective
slopes change).%
% Recent work suggests a clumping of experimental values around the
% mean-field and front-propagation predictions~\cite{DuriZapp00}.
  (A more systematic and critical review of the exponent 
  measurements in the literature may be found in Table I of 
  reference~\cite{DuriZapp00}. 
  One may also find in section V.A of their paper a discussion of the 
  relation between the experiments and the theoretical universality classes.)
} 
\label{fig:ExponentExpt}
\end{figure}

There are several basic critical exponents, which arise in many different
combinations depending on the physical property being studied. 
We've seen that at $R_c$ and near $H_c$ the probability of having an
avalanche of size $S$ goes as $S^{-\tau}$.%
  \footnote{We will see that when the entire hysteresis loop is considered 
  this changes to $S^{-(\tau+\sigma\beta\delta)}$.}
The cutoff in the avalanche size distribution in figure~\ref{fig:AvalHisto}
gets larger as one approaches the critical disorder as $(R-R_c)^{-\sigma}$
(figure~\ref{fig:AvalHisto}). The typical length of the largest
avalanche goes as $(R-R_c)^{-\nu}$. The jump in the magnetization 
(figure~\ref{fig:BetaDeltaJump}) goes as $(R-R_c)^\beta$, and at $R_c$ the
magnetization $(M-M_c)\sim (H-H_c)^{1/\delta}$.%
  \footnote{Don't confuse the small change in scale $\delta$ earlier with
  the critical exponent $\delta$ here.} 
Finally, we need to know how time rescales: the duration of an avalanche
of spatial extent $L$ typically will go as $L^z$. 

Any physical property that shows singular behavior at the critical point
will have a critical exponent that can be written in terms of these basic
ones (figure~\ref{fig:ExponentExpt}).%
  \footnote{Indeed, there are relations even between these exponents: see
  section~\ref{sec:ExponentRelations}.}

\begin{figure}[thb]
  \begin{center}
    \epsfxsize=8cm
    \epsffile{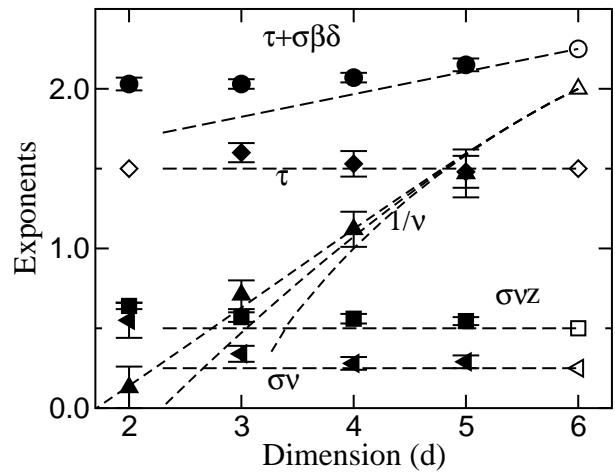}
  \end{center}
\caption{{\bf Universal Critical Exponents in Various Spatial 
Dimensions.}~\cite{SethDahmMyer01}
We test our $\epsilon$-expansion predictions~\cite{DahmSeth96} by
measuring the various critical exponents numerically 
in up to five spatial dimensions~\cite{PerkDahmSeth95,PerkDahmSeth99}.
The various exponents are
described in the text. All of the exponents are calculated only to
linear order in $\epsilon$, except for the correlation length exponent $\nu$,
where we use results from other models. The agreement even in three
dimensions is remarkably good, considering that we're expanding in $6-D$ where
$D=3$! 
We should note that perturbing in dimension for our system is not
only complicated, but also controversial.
Our expansion uses the Martin Siggia Rose 
formalism~\cite{MartSiggRose73,BausJansWagn76,DeDo78,SompZipp82,Zipp84,NattStep+92,NaraFish93,Nara00} to describe a deterministic dynamical system without 
any temperature fluctuations, while the calculation for the equilibrium
model involves temperature fluctuations and no history dependence at all.
Even so, we have shown
that our $6-\epsilon$ expansion for the critical exponents of the 
{\it zero-temperature, nonequilibrium} model maps to all orders in 
$\epsilon$ onto that of the equilibrium model.
The $6-\epsilon$ expansion for the equilibrium model, however, has been 
controversial for decades~\cite{Feld02}.
Recently the thermal $6-\epsilon$ expansion of the {\it equilibrium}
model has been called into question~\cite{Feld02,BrezdeDo98,BrezdeDo99}
not just due to non perturbative corrections, as was previously assumed, 
but due to previously neglected higher order terms in the expansion. 
The implications of this controversy for our non--equilibrium
renormalization group treatments is not yet known.
} 
\label{fig:ExponentDim}
\end{figure}

To specialists in critical phenomena, these exponents are central; whole
conversations will seem to rotate around various combinations of Greek
letters. Critical exponents are one of the relatively easy things to
calculate from the various analytic approaches, and so have attracted
the most attention. They are derived from the eigenvalues of the
linearized flows about the fixed point $\mathbf{S^*}$ in
figure~\ref{fig:RGFlows}. Figure~\ref{fig:ExponentDim} shows
our numerical estimates for several critical exponents in our model in
various spatial dimensions, together with our $6-\epsilon$ expansions for
them. Of course the key challenge is not to get analytical work to agree
with numerics: it's to get theory to agree with experiment. 
Figure~\ref{fig:ExponentExpt}
shows that our model does rather well in describing a wide variety of
experiments, but that the two rival models (with different flows around
their fixed points) also fit.

\begin{figure}[thb]
  \begin{center}
    \epsfxsize=8cm
    \epsffile{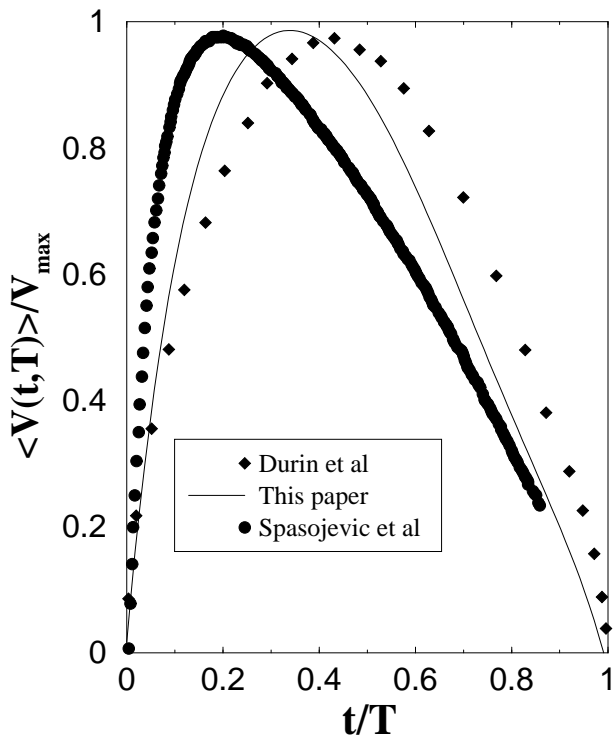}
  \end{center}
\caption{{\bf Comparison of experimental average pulse shapes} for fixed pulse
duration, as measured by three different 
groups~\cite{MehtMillDahm+02,DuriZapp00,DuriZapp01,SpasBukvMilo+96}.
Our theories don't only predict power laws: they should describe all 
behavior on long length and time scales (at least in a statistical sense).
In particular, by fixing parameters one can predict what are called 
{\em scaling functions}. If we average the voltage as a function of time 
over all avalanches of a fixed duration, we get an average shape. 
In these three experiments, they find that this shape is the same for 
different durations: by rescaling the time by the duration $T$ and the
voltage by the maximum of the average curve, the curves within a given
experiment collapse onto one. The three experiments, however, do not all
collapse onto the same curve. This could mean that they are in different
universality classes, or that we don't understand this dynamical scaling
completely.
} 
\label{fig:AmitExptPulseShapeCollapse}
\end{figure}

\begin{figure}[thb]
  \begin{center}
    \epsfxsize=8cm
    \epsffile{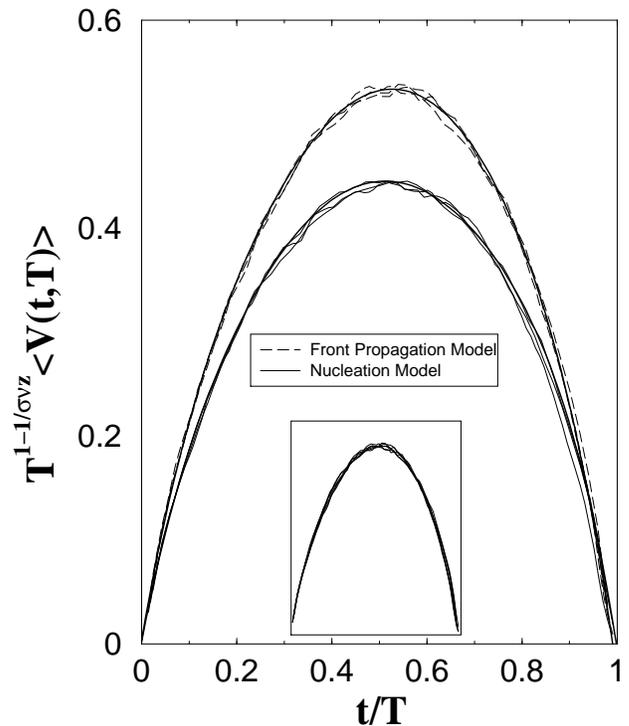}
  \end{center}
\caption{{\bf Comparison of theoretical average pulse shape scaling functions}
for our nucleated model and the front propagation model~\cite{MehtMillDahm+02}.
The front propagation models have $1/\sigma\nu z = 1.72\pm0.03$ in this 
collapse; our nucleation model has $1/\sigma\nu z = 1.75\pm 0.03$ (in
principle, there is no reason to believe these two should agree). The
inset shows the two curves rescaled to the same height (the overall
height is a non--universal feature): they are quantitatively different, 
but far more similar to one another than either is to the experimental
curves in figure~\ref{fig:AmitExptPulseShapeCollapse}. The mean--field
model apparently has a scaling function which is a perfect inverted 
parabola~\cite{Kunt01}. The ABBM model, which is also mean--field like, 
interestingly has a different shape, that of one lobe of a
sinusoid~\cite[eq. 102]{DuriZapp04}.
} 
\label{fig:AmitTheoryPulseShapeCollapse}
\end{figure}

Critical exponents are not the be-all and end-all: many other scaling
predictions, explaining wide varieties of behavior, are quite easy to
extract from numerical simulations. Universality extends even to those
long length scale properties for which one cannot write formulas.
Perhaps the most important of these other predictions are the universal
scaling functions. For example, lets consider the time history of the
avalanches, $V(t)$, denoting the number of domains flipping per unit time.
(We call it $V$ because it's usually measured as a voltage in a pickup
coil.) Each avalanche has large fluctuations, but one can average over
many avalanches to get a typical shape.  
Figures~\ref{fig:AmitExptPulseShapeCollapse} 
and~\ref{fig:AmitTheoryPulseShapeCollapse}
show averages over all avalanches of fixed duration $T$. Lets call this 
$\langle V\rangle(T,t)$.
Universality again suggests that this average should be the same for
experiment and a successful theory, apart from an overall shift in time
and voltage scales: 
\begin{equation}
\langle V\rangle_{\mathrm{exp}}(T,t) 
	= A \langle V\rangle_{\mathrm{th}}(T/B, t/B).
\end{equation}
Comparing our model to itself with a shifted time scale becomes simple
if we change variables: let $v(T,t/T) = \langle V\rangle(T,t)$, so 
$v(T,t/T) = A v(T/B,t/T)$. Here $t/T$ is a particularly simple example of a 
{\em scaling variable}. Now, if we rescale time by a small factor 
$B=1/(1-\delta)$, we have
$v(T, t/T) = (1 + b) v(t/T, (1-\delta) T)$. Again, making $\delta$ small we
find $b v = T \partial v/\partial T$, with solution $v = v_0 T^b$. However,
the integration constant $v_0$ will now depend on $t/T$, $v0 =V(t/T)$, 
so we arrive at the scaling form
\begin{equation}
\langle V\rangle (t,T) = T^b V(t/T),
\end{equation}
where the entire scaling function $V$ is a universal prediction of the theory. 

Figures~\ref{fig:AmitExptPulseShapeCollapse} 
and~\ref{fig:AmitTheoryPulseShapeCollapse}
show the universal scaling functions $V$ for two models
and three experiments. For our model, we've drawn what are called scaling
collapses, a simple but powerful way to both check that were in the
scaling regime, and to measure the universal scaling function. Using the
form of the scaling equation Eq.[2], we simply plot 
$T^b \langle V\rangle(t,T)$ versus $t/T$, for a series of long times $T$.
All the plots fall onto the same
curve. This tells us that our avalanches are large enough to be
self-similar. (If in your scaling collapse the corresponding plots do
not all look alike, then any power laws you have measured are probably
accidental.) It also provides us with a numerical evaluation of the
scaling function $V$. Note that we use $1/\sigma\nu z-1$ for the critical
exponent $b$. This is an example of an exponent equality: easily derived
from the fact that 
\begin{equation}
\langle S\rangle(T) = \int\langle V\rangle(t,T)\, dt
	=  \int T^b V(t/T) dt \sim T^{b+1},
\end{equation}
and the scaling
relation $\langle S\rangle(T) \sim T^{-1/\sigma\nu z}$.

Notice that the two models and the three experiments have quite different
shapes for $V$. How do we react to this? Our
models are falsified if any of the predictions are shown to be wrong
asymptotically on long length and time scales: hence our theory is 
either wrong (inapplicable to these systems) or somehow at least incomplete. 
Incorporating
insights from careful experiments to refine the theoretical models has
historically been crucial in the broad field of critical phenomena. The
message we emphasize here is that scaling functions can provide a
sharper tool for discriminating between different universality classes
than critical exponents.

Broadly speaking, most common properties that involve large length and
time scales have scaling forms: using self-similarity, one can write
functions of $N$ variables in terms of scaling functions of $N-1$ variables:
$F(x,y,z) = z^{-\alpha} F(x/z^{\beta}, y/z^{\gamma})$. In the inset to 
figure~\ref{fig:AvalHisto}, we show the scaling
collapse for the integrated avalanche size distribution (derived as
equation~\ref{eq:curlyD}): 
\begin{equation}
D_{\mathrm{int}}(S,R) = 
  S^{-(\tau+\sigma\beta\delta)}{\cal D}_{\mathrm{int}}((R-R_c)/ S^{-\sigma}).
\end{equation}
This example illustrates that scaling works not only at $R_c$ but also
near $R_c$; the unstable manifold in figure~\ref{fig:RGFlows} governs
the behavior for systems near the critical manifold $C$.

\section{Measuring Exponents and Scaling Functions}

The simulations provide a rich variety of measured quantities, each with
a characteristic universal scaling form and associated critical exponents.
In this section we focus on a few key ones of particular experimental
significance.%
  \footnote{This section follows closely the presentation 
  in~\cite{PerkDahmSeth99}.}

We'll discuss the following properties obtained from the simulation%
\begin{itemize}
\item
the magnetization $M(H,R)$ as a function of the external field $H$.
\item
the avalanche size distribution integrated over the field $H$, $D_{int}(S,R)$. 
\item
the avalanche correlation function integrated over the field $H$,
$G_{int}(x,R)$. 
\item
the distribution of avalanche durations $D_{t}^{(int)}(S,t)$
as a function of the avalanche size $S$, at
$R=R_c$, integrated over the field $H$.
\item
the energy spectrum $E(\omega)$ of the Barkhausen noise for
various models at criticality.
\end{itemize}

\subsection{Magnetization Curves}

Unfortunately the most obvious measured quantity in our simulations,
the magnetization curve $M(H)$, is the one which collapses least well in our
simulations. We start with it nonetheless.

\begin{figure}[thb]
  \begin{center}
    \epsfxsize=8cm
    \epsffile{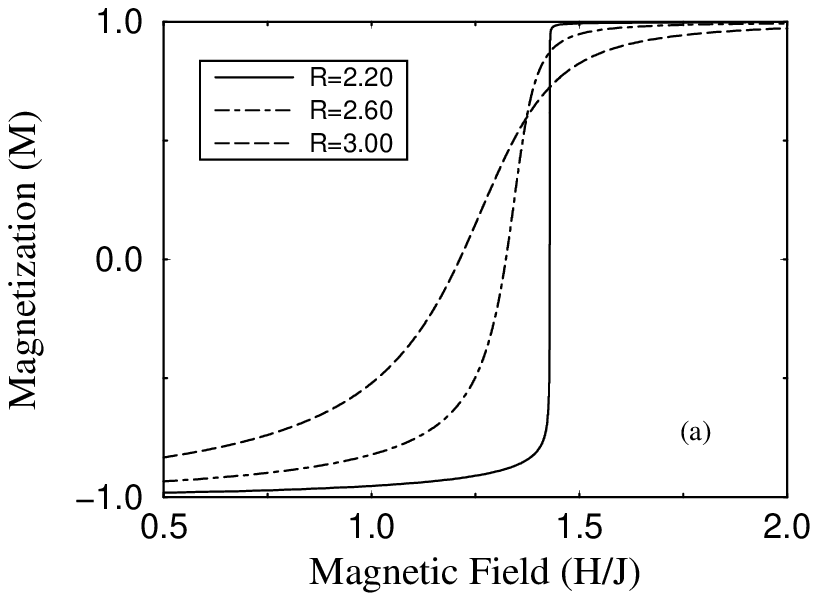}
    \epsfxsize=8cm
    \epsffile{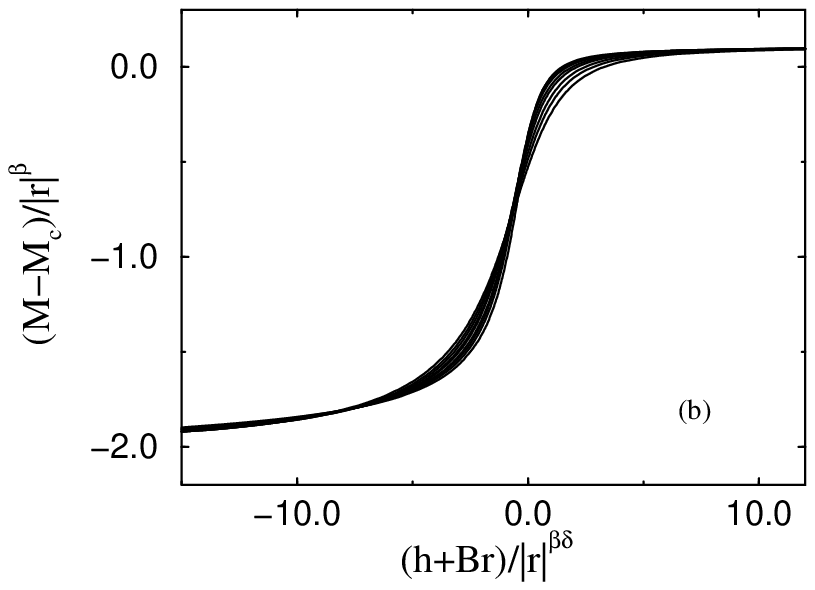}
  \end{center}
\caption{{\bf $M(H)$ curves at different disorders, and (poor) universal
scaling collapse.}~\cite{PerkDahmSeth99}
} 
\label{fig:MofH}
\end{figure}

The top figure \ref{fig:MofH} shows the magnetization curves obtained from
our simulation in $3$ dimensions for several values of the disorder $R$.
As the disorder $R$ is decreased, a discontinuity or jump in the
magnetization curve appears where a single avalanche occupies a large
fraction of the total system. In the thermodynamic limit this would be
the infinite avalanche: the largest disorder at which it occurs is the
critical disorder $R_c$. For finite size systems, like the ones we use
in our simulation, we observe an avalanche which spans the system at a
higher disorder, which gradually approaches $R_c$ as the system size
grows.

\begin{figure}[thb]
  \begin{center}
    \epsfxsize=8cm
    \epsffile{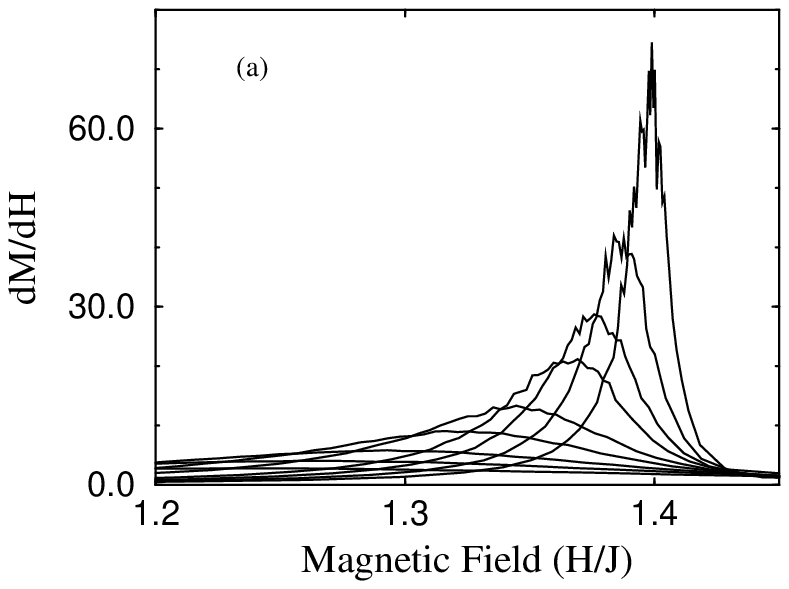}
    \epsfxsize=8cm
    \epsffile{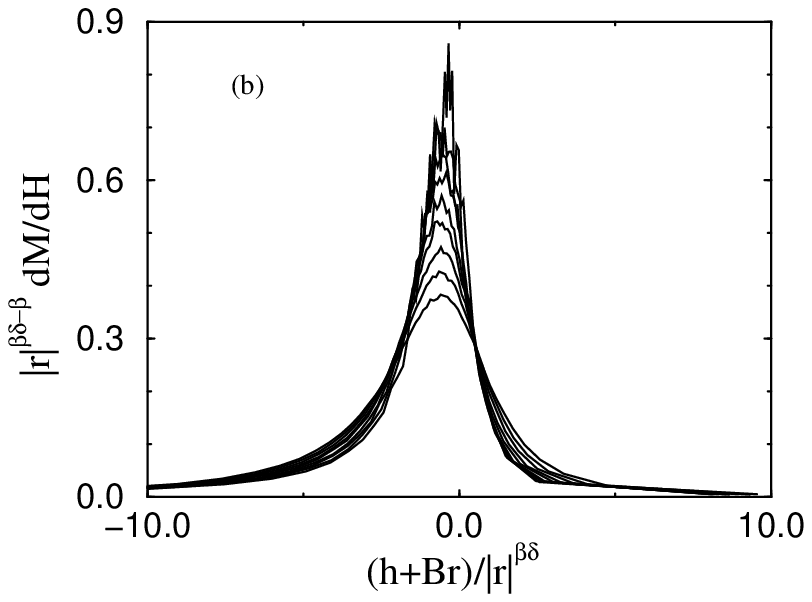}
  \end{center}
\caption{{\bf $\frac{dM}{dH}(H)$ curves at different disorders, and 
(poor) universal scaling collapse.}~\cite{PerkDahmSeth99}
While the curves are not collapsing onto a single curve, the quality of the
collapse is quite similar to that found at similar distances from $R_c$
in mean field theory~\protect\cite{PerkDahmSeth96}, for which we know 
analytically that scaling works as $R \to R_c$.
} 
\label{fig:dMdH}
\end{figure}

Figure~\ref{fig:dMdH} shows the slope $dM/dH$ and its scaling collapse.
By using this derivative, the critical region is emphasized as the peak
in the curve, and the dependence on the parameter $M_c$ drops out.
The lower graphs in figure~\ref{fig:MofH} and~\ref{fig:dMdH}
show the scaling collapses of the magnetization and its slope. Clearly
in neither case is all the data collapsing onto a single curve. This
would be distressing, were it not for the fact that this also occurs in
mean field theory~\cite{PerkDahmSeth96} at a similar distance to the critical
point.

Because the scaling of the magnetization is so bad, we use other quantities
to estimate the critical exponents and the location of the critical point
(Tables~I and~III). Fixing these quantities,
we use the collapse of the $dM/dH$ curves to extract the rotation $B$ mixing
the experimental variables $r$ and $h$ into the scaling variable $h'=h + B r$
(equation~\ref{eq:scaling_2} and following discussion). 
Recent work suggests that much of the difficulty in collapsing
the magnetization curves is surmountable by adding analytical corrections to 
scaling~\cite{SethUnp}.

\subsection{Avalanche Size Distribution}
\label{sbsec:AvalHisto}

In our model the spins flip in avalanches: each spin can kick over one or
more neighbors in a cascade.  These
avalanches come in different sizes. The integrated avalanche size
distribution is the size distribution of all the avalanches that occur
in one branch of the hysteresis loop (for $H$ from $-\infty$ to
$\infty$). Figure~\ref{fig:AvalHisto}~\cite{PerkDahmSeth95} shows some of the
raw data (thick lines) in $3$ dimensions.   Note that the curves follow an
approximate power law behavior over several decades. Even $50\%$ away from
criticality (at $R=3.2$), there are still two decades of scaling, which
implies that the critical region is large. In experiments, a few decades
of scaling could be interpreted in terms of self-organized criticality.
However, our model and simulation suggest that several decades of
power law scaling can still be present rather {\em far} from the
critical point (note that the size of the critical region is
non--universal). The slope of the log-log avalanche size distribution at
$R_c$ gives the critical exponent $\tau + \sigma \beta \delta$. Notice,
however, that the apparent slopes in figure~\ref{fig:AvalHisto} continue to
change even after several decades of apparent scaling is obtained.
The cutoff in the power law diverges as the critical disorder $R_c$ is
approached. This cutoff size scales as $S \sim |r|^{-1/\sigma}$.

These critical exponents can be obtained by using a scaling collapse for
the curves of figure~\ref{fig:AvalHisto}, shown in the inset. 
The scaling form for the avalanche sizes as a function of $R$ and $H$ is
\begin{equation}
D(S,R,H) \sim S^{-\tau} {\cal D}(S^\sigma |r|, h/r^{\beta\delta}).
\end{equation}
We can find the scaling form for the distribution of avalanche sizes
integrated over $H$ by integrating this formula, changing variables to
$y=h/r^{\beta\delta}$, and rewriting the resulting formula in terms of
$S$ to a power times a function of $S^\sigma r$. For $R>R_c$,
\begin{eqnarray}
\label{eq:curlyD}
D_{int}(S,R) 
	&=& \int D(S,R,H) \, dH \\
	&\propto& \int S^{-\tau}{\cal D}(S^\sigma r,h/r^{\beta\delta}) \, dh \cr
	&=& S^{-\tau} r^{\beta\delta} \int {\cal D}(S^\sigma r,y) \, dy \cr
	&=& S^{-(\tau+\sigma\beta\delta)} (S^\sigma r)^{\beta\delta} 
		\int {\cal D}(S^{\sigma}r,y) \, dy \cr
	&=& S^{-(\tau+\sigma\beta\delta)} {\cal D}^{(int)}_{+} (S^{\sigma} r) 
		\nonumber
\end{eqnarray}
where ${\cal D}_{+}^{(int)}$ is the scaling function for the integrated
avalanche size distribution (the $+$ sign
indicates that the collapsed curves are for $R > R_c$). We are sufficiently
far from the critical point that corrections to scaling are important:
as described in reference~\cite{PerkDahmSeth96}, we do collapses for small
ranges
of $R$ and then linearly extrapolate the best--fit critical exponents to $R_c$.
We estimate from this curve that the critical
exponents $\tau+\sigma\beta\delta=2.03$ and $\sigma=0.24$

\begin{figure}[thb]
  \begin{center}
    \epsfxsize=8cm
    \epsffile{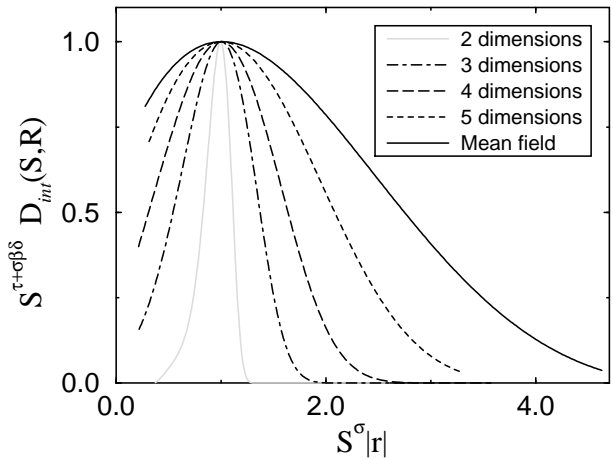}
  \end{center}
\caption{{\bf Universal scaling functions ${\cal D}_{int}(S^\sigma |r|)$ 
for the avalanche size distributions in different spatial 
dimensions.}~\cite{PerkDahmSeth99} The inset of figure~\ref{fig:AvalHisto} 
shows the scaling collapse for three dimensions.
} 
\label{fig:curlyDvsDim}
\end{figure}

The scaling function ${\cal D}_{+}^{(int)}(X)$ with $X=S^{\sigma}|r|$
is a universal prediction of our model. To facilitate comparisons with
experiments, we fit a curve to the data collapse in the inset of
figure~\ref{fig:AvalHisto}. We have fit the scaling collapses in dimensions
$3$, $4$, and $5$ to a phenomenological form of an exponential times
a polynomial. In three dimensions, our fit is 
\begin{eqnarray}
&& {\cal D}_{+}^{\it (int)}(X)\ =\ e^{-0.789X^{1/\sigma}}\ \times \\
&&~~~~ (0.021+0.002X+0.531X^2-0.266X^3+0.261X^4) \nonumber
\label{eq:aval_fit_3d}
\end{eqnarray}
where $1/\sigma=4.20$.  The
distribution curves obtained using the above fit are plotted (thin lines
in figure~\ref{fig:AvalHisto}) alongside the raw data (thick lines). They
agree remarkably well even far above $R_c$. We should recall though,
that the fitted curve to the collapsed data can differ from the ``real''
scaling function even for large sizes and close to the critical disorder (in
mean field~\cite{PerkDahmSeth96} the error in the corresponding curve was about
$10\%$).
The scaling function in the inset of figure~\ref{fig:AvalHisto} has a
peculiar shape: it grows by a factor of ten before cutting off. The
consequence of this bump in the shape is that in the simulations it takes many
decades in the size distribution for the slope to converge to the
asymptotic power law. This can be seen from the comparison between a
straight line fit through the $R=2.25$ (billion spin) simulation in 
figure~\ref{fig:AvalHisto} and the asymptotic power law $S^{-2.03}$ obtained
from extrapolating the scaling collapses (thick dashed straight line in
the same figure). A similar bump exists in other dimensions and mean
field as well. Figure~\ref{fig:curlyDvsDim} shows the scaling functions in
different dimensions and in mean field. In this graph, the scaling
functions are normalized to one and the peaks are aligned (the scaling
forms allow this). The curves plotted in figure~\ref{fig:curlyDvsDim} are
not raw data but fits to the scaling collapse in each dimensions, as was
done in the inset of figure~\ref{fig:AvalHisto}. 

It is clear from the figure that the growing bump in the scaling curves
as the dimension decreases is a foreshadowing of a zero in the scaling
curve in two dimensions.

\subsection{Avalanche Correlations}

\begin{figure}[thb]
  \begin{center}
    \epsfxsize=8cm
    \epsffile{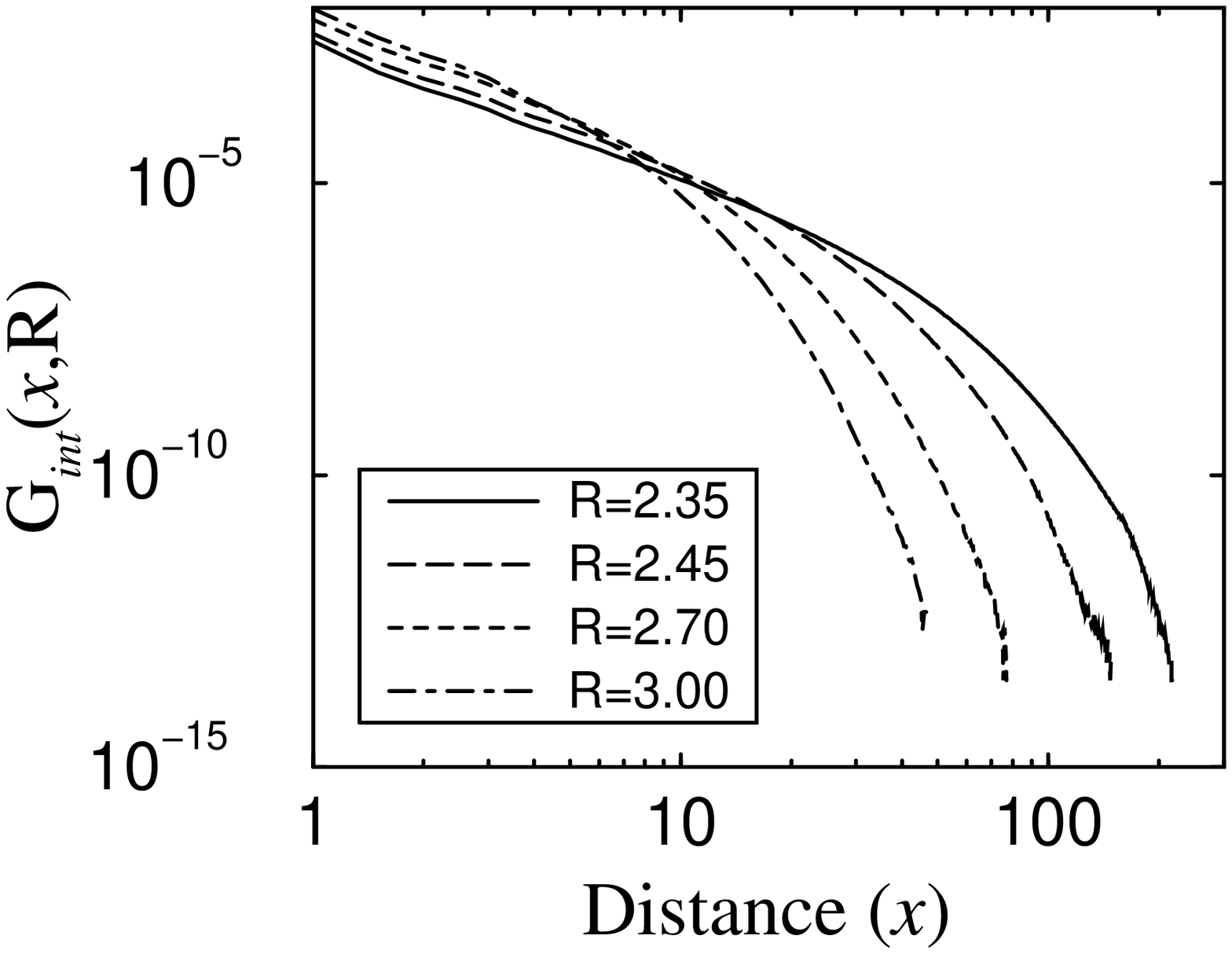}
    \epsfxsize=8cm
    \epsffile{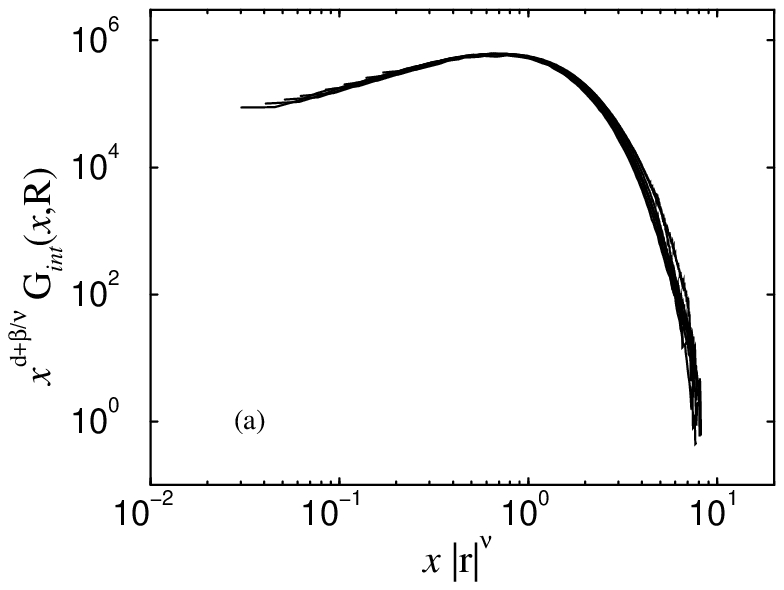}
  \end{center}
\caption{{\bf Avalanche correlation function} integrated over field $H$
at various disorders, with scaling collapse~\cite{PerkDahmSeth99}.
} 
\label{fig:Correl}
\end{figure}

The avalanche correlation function $G(x,R,H)$ measures the probability
that the initial spin of an avalanche will trigger, in that avalanche,
another spin a distance $x$ away. From the renormalization group
description~\cite{DahmSeth93,DahmSeth96}, close to the critical point and for
large distances $x$, the correlation function is given by:
\begin{equation}
G(x,R,H) \sim {1 \over {x^{d-2+\eta}}}\ {\cal G}_{\pm}(x/{\xi(r,h)})
\label{eq:correl_1}
\end{equation}
where $r$ and $h$ are respectively the reduced disorder and field,
${\cal G_{\pm}}$ ($\pm$ indicates the sign of $r$) is the scaling
function, $d$ is the dimension, $\xi$ is the correlation length, and
$\eta$ is called the ``anomalous dimension''. Corrections can be shown
to be subdominant (appendix~\ref{app:ScalingForms}). The correlation length
$\xi (r,h)$ is a macroscopic length scale in the system which is 
on the order of the mean linear extent of the largest avalanches. 
At the critical field $H_c$ (h=0) and near
$R_c$, the correlation length scales like $\xi \sim |r|^{-\nu}$, while
for small field $h$ it is given by 
\begin{equation}
\xi \sim |r|^{-\nu}\ {\cal Y}_{\pm}(h/|r|^{\beta\delta})
\label{eq:correl_1b}
\end{equation}
where ${\cal Y}_{\pm}$ is a universal scaling function. The avalanche
correlation function should not be confused with the cluster or
``spin-spin'' correlation which measures the probability that two spins
a distance $x$ away have the same value. (The algebraic decay for this
other, spin-spin correlation function at the critical point ($r=0$ and
$h=0$), is $1/{x^{d-4+{\tilde \eta}}}$~\cite{DahmSeth93}.)

We've mostly used, for historical reasons, a slightly different
avalanche correlation function, which scales the same way as the
``triggered'' correlation function $G$ described above. Our function
basically ignores the difference between the triggering spin and the
other spins in the avalanche: alternatively, it calculates for
avalanches of size $S$ the correlation function for pairs of spins, and
then averages over all avalanches (weighting each avalanche equally).
We've checked that our correlation function agrees to within 3\% with
the ``triggered'' correlation function described above, for $R>R_c$ in
three dimensions and above. (In two dimensions, the two definitions
differ more substantially, but appear to scale in the same
way.)

We have measured the avalanche correlation function integrated over the
field $H$, for $R>R_c$. For every avalanche that occurs between
$H=-\infty$ and $H =+\infty$, we keep a count on the number of times a
distance $x$ occurs in the avalanche. 
The spanning avalanches are not included in
our correlation measurement. Figure~\ref{fig:Correl} shows several
avalanche correlation curves in $3$ dimensions for $L=320$. The scaling
form for the avalanche correlation function integrated over the field
$H$, close to the critical point and for large distances $x$, is
obtained by integrating equation (\ref{eq:correl_1}):
\begin{equation}
G_{\it int}(x,R) \sim \int {1 \over {x^{d-2+\eta}}}\
{\cal G}_{\pm}\Bigl(x/{\xi(r,h)}\Bigr)\ dh
\label{eq:correl_2}
\end{equation}
Using equation (\ref{eq:correl_1b}) and defining $u=h/|r|^{\beta\delta}$,
equation (\ref{eq:correl_2}) becomes:
\begin{equation}
G_{\it int}(x,R) \sim |r|^{\beta\delta}{x^{-(d-2+\eta)}} \int
{\cal G}_{\pm}\Bigl(x/|r|^{-\nu} {\cal Y}_{\pm} (u)\Bigr)\ du
\label{eq:correl_3}
\end{equation}
The integral ($\cal I$) in equation (\ref{eq:correl_3}) is a function of
$x|r|^{\nu}$ and can be written as:
\begin{equation}
{\cal I} =
(x|r|^{\nu})^{-\beta\delta/\nu}\ {\widetilde {\cal G}}_{\pm}(x|r|^{\nu})
\label{eq:correl_4}
\end{equation}
to obtain the scaling form:
\begin{equation}
G_{\it int}(x,R) \sim {1 \over x^{d+\beta/\nu}}\
{\widetilde {\cal G}}_{\pm}(x|r|^{\nu})
\label{eq:correl_5}\end{equation}
where we have used the scaling relation $(2-\eta)\nu=\beta\delta-\beta$
(see~\cite{DahmSeth93} for the derivation).

The bottom figure~\ref{fig:Correl} shows the integrated avalanche
correlation curves collapse in $3$ dimensions for $L=320$ and $R>R_c$.
The exponent $\nu$ is obtained from such collapses by extrapolating to
$R = R_c$ as was done for other collapses~\cite{PerkDahmSeth96}. The exponent
$\beta/\nu$ can be obtained from these collapses too, but we found it
better to use the jump in magnetization~\cite{PerkDahmSeth99}, which
near the critical point involves several spanning avalanches.
Recent work~\cite{PereVive03,VivePere04,IllaVive04} suggests that the jump in 
magnetization
may be dominated by only one of these spanning avalanches, and their
work suggests that there may be two exponents related to our $\beta$, one
substantially larger (their $\beta_c\sim 0.15\pm0.08$).
The value of $\beta/\nu$ listed in Table~I is derived
exclusively from the magnetization discontinuity collapses.

\begin{figure}
  \begin{center}
    \epsfxsize=8cm
    \epsffile{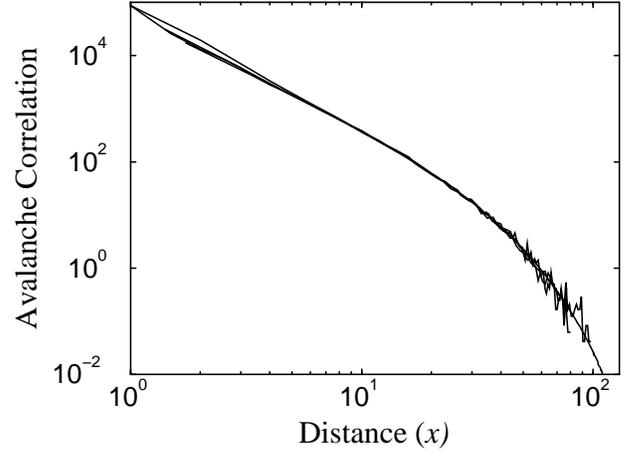}
  \end{center}
\caption{{\bf Anisotropies in the avalanche correlation function}.
Four curves are
shown on the graph: one is the avalanche correlation function integrated
over the field $H$ (as in figure~\protect\ref{fig:Correl}), while the
other three are measurements of the correlation along the three axis,
the six face diagonals, and the four body diagonals. Avalanches
involving more than four spins show no noticeable anisotropy: the
critical point appears to have spherical symmetry. The same result is
found in $2$ dimensions.}
\label{fig:CorrelAnisotropy}
\end{figure}

One of the most artificial features of our model is that the domains
are arranged in a regular grid. It is thus important to check that this
grid is not important for our scaling properties: if the avalanches looked
noticeably cubical, we would have problems.
We have looked for possible anisotropies in the integrated
avalanche correlation function in $2$ and $3$ dimensions. The
anisotropic integrated avalanche correlation functions are measured
along ``generalized diagonals'': one along the three axis, the second
along the six face diagonals, and the third along the four body
diagonals. We compare the integrated avalanche correlation function and
the anisotropic integrated avalanche correlation functions to each
other, and find no anisotropies in the correlation, as can be seen from
figure~\ref{fig:CorrelAnisotropy}. Spherical symmetry emerges at the
critical point on long length and time scales.

\subsection{Avalanche Duration Measurement}

The scaling relation between the duration $T$ of an avalanche (the time 
it takes to occur) and the linear size $\xi$ of the avalanche 
defines the critical exponent $z$
\begin{equation}
T \sim \xi^z
\label{eq:time_1}
\end{equation}
The exponent $z$ is known as the dynamical critical exponent. 
Equation~(\ref{eq:time_1}) gives the scaling for the time it takes for a
spin to ``feel'' the effect of another a distance $\xi$ away. Since the
correlation length $\xi$ scales like $r^{-\nu}$ close to the critical
disorder, and the characteristic size $S$ as $r^{-1/\sigma}$, the duration
$T$ then scales with avalanche size as:
\begin{equation}
\label{eq:TvsS}
T \sim S^{\sigma \nu z}
\label{eq:time_2}
\end{equation}

In our simulation, we measure the distribution of durations for each
avalanche size $S$. The probability $D_t(S,R,H,T)$ that an avalanche of 
size $S$ will be of duration $T$ close to the critical field $H_c$ and
critical disorder $R_c$ has the scaling form
\begin{equation}
D_t(S,R,H,T) \sim S^{-q}\ {\cal D}_{\pm}^{(t)} (S^{\sigma}|r|,
h/|r|^{\beta\delta}, T/S^{\sigma\nu z})
\label{eq:time_3}
\end{equation}
where $q=\tau +\sigma\nu z$, and is defined such that
\begin{eqnarray}
\int_{-\infty}^{+\infty} \!\! \int_{1}^{\infty} D_t(S,R,H,T)\ dH\ dt\ =\
\nonumber \\
S^{-(\tau + \sigma\beta\delta)}\ {\cal D}_{\pm}^{(int)}(S^{\sigma}|r|)
\label{eq:time_4}
\end{eqnarray}
where ${\cal D}_{\pm}^{(int)}$ was defined in the integrated
avalanche size distribution section. The avalanche time distribution
integrated over the field $H$, at the critical disorder ($r=0$) is:
\begin{equation}
D_t^{(int)}(S,T)\ \sim\
t^{-{(\tau + \sigma\beta\delta + \sigma\nu z) /\sigma\nu z}}\ 
{\cal D}_t^{(int)}(T/S^{\sigma\nu z})
\label{eq:time_5}
\end{equation}
as obtained from equation (\ref{eq:time_3}) (reference~\cite{PerkDahmSeth96}).

\begin{figure}[thb]
  \begin{center}
    \epsfxsize=8cm
    \epsffile{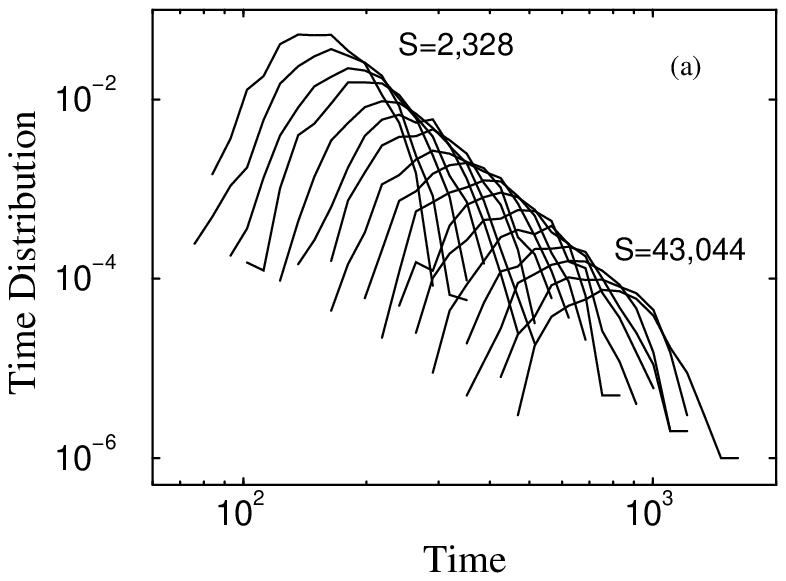}
    \epsfxsize=8cm
    \epsffile{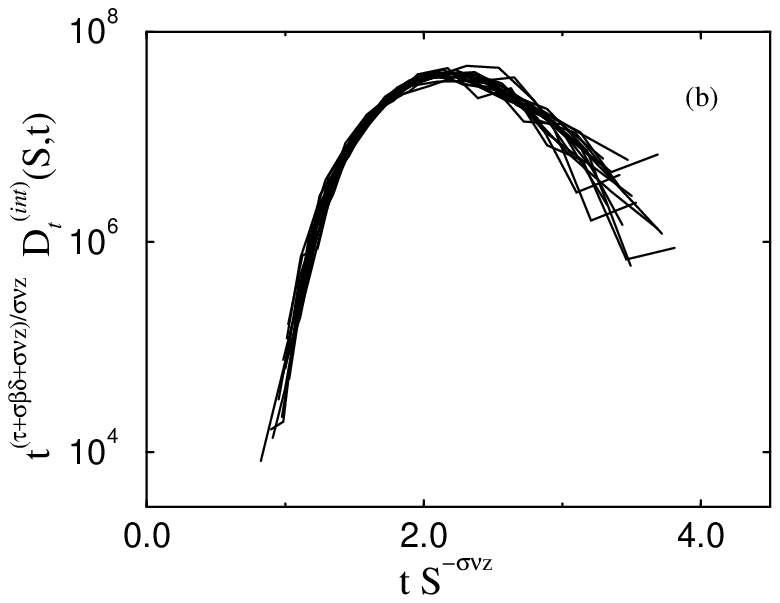}
  \end{center}
\caption{{\bf Distribution of avalanche durations for avalanches of fixed size
$S$, for various sizes $S$, and their scaling collapse.}~\cite{PerkDahmSeth99}
} 
\label{fig:AvalTime}
\end{figure}

Figures~\ref{fig:AvalTime} show the avalanche
time distribution integrated over the field $H$ for different avalanche
sizes, and a collapse of these curves using the above scaling form.

\subsection{Energy Spectrum}

\begin{figure}[thb]
  \begin{center}
    \epsfxsize=8cm
    \epsffile{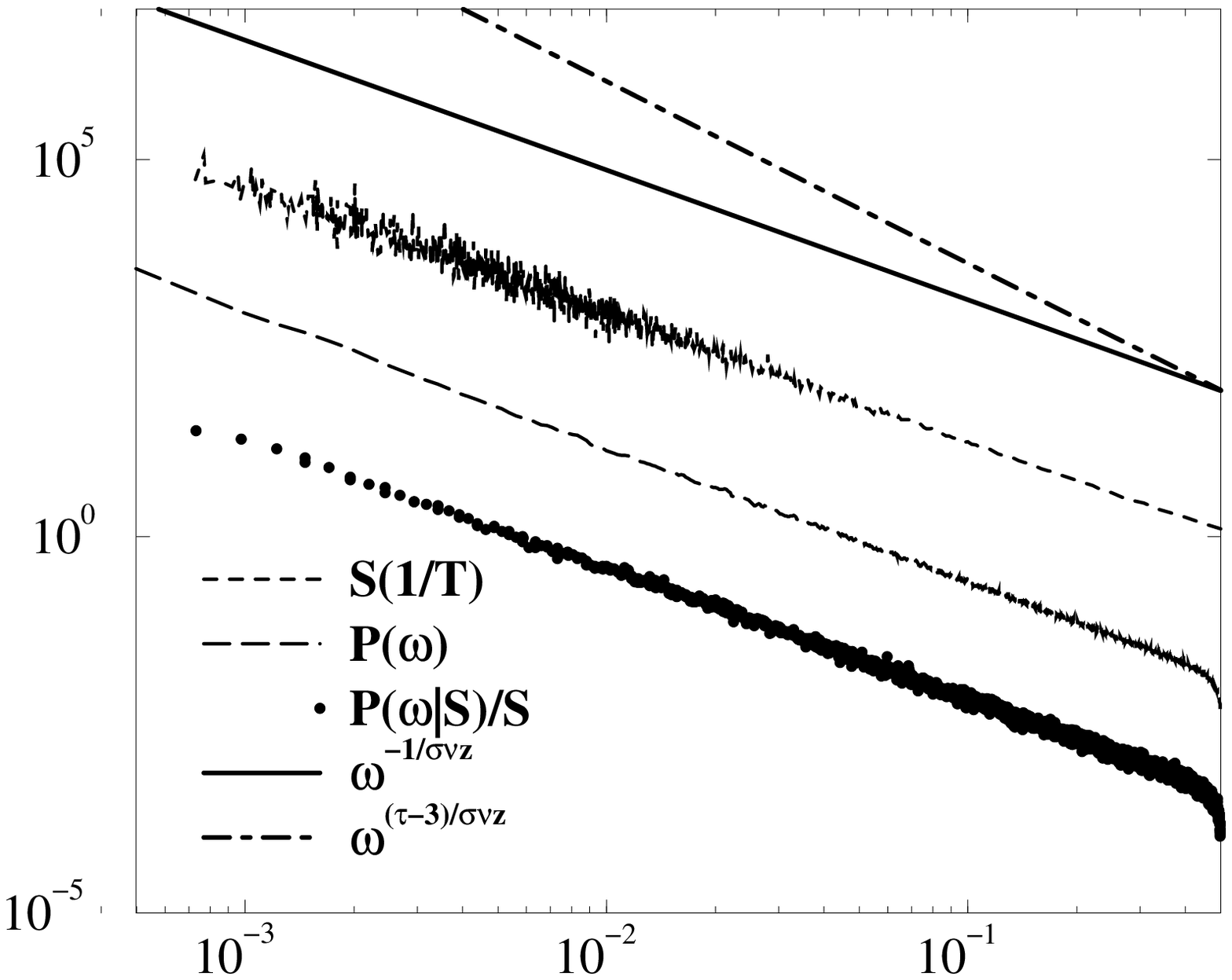}
    \epsfxsize=8cm
    \epsffile{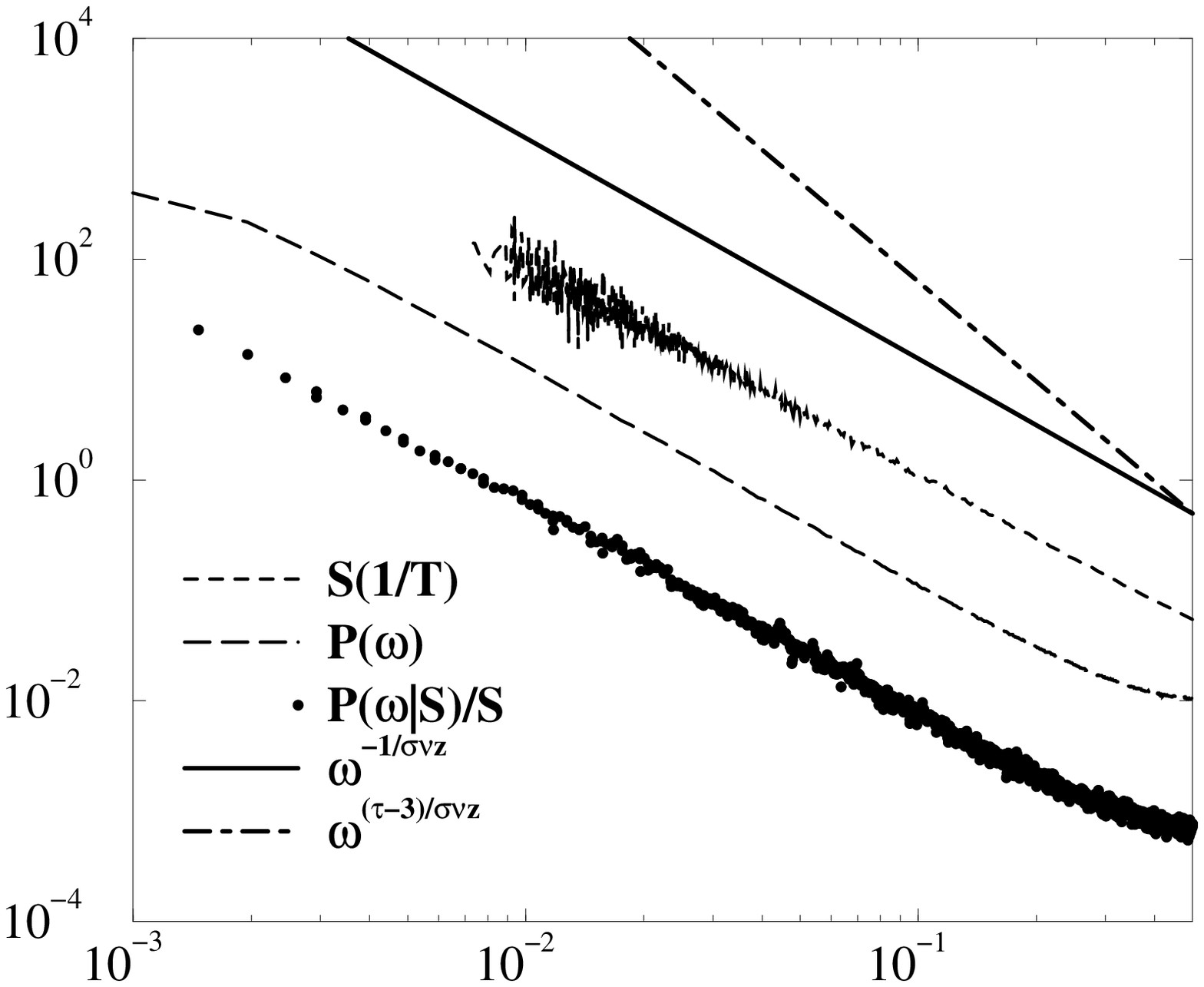}
  \end{center}
\caption{{\bf Power spectrum scaling and relatives.} Top: short--range
model; bottom: mean field (long--range dipolar forces~\cite{KuntSeth00}.)
} 
\label{fig:PowerScaling}
\end{figure}

The energy spectrum of the voltage $V(t)$ as a function of time for our model
(given by adding up the the voltage traces of each avalanche, as in 
figure~\ref{fig:AvalancheTypicalShape}), 
\begin{equation}
E(\omega) = \int e^{-i\omega \tau} \langle V(t) V(t+\tau) \rangle
	= |\widetilde V(\omega)|^2
\end{equation}
is a commonly measured experimental quantity.%
  \footnote{The power spectrum is $E(\omega)$ divided by the integration
  time.}
There are two distinct
contributions to this power spectrum: (1)~at high frequencies, the
incoherent sum of the power spectra $E_{\mathrm{inc}}$ of the individual
avalanches, and (2)~at 
low frequencies, a term representing the correlations between avalanches.
As the system is forced more and more slowly, the internal dynamics of the
avalanches is unchanged but their separation in time increases, so these
two contributions separate. Typical experiments seem to be dominated by
the contribution~(1) from the internal dynamics within an 
avalanche~\cite{KuntSeth00}. In this section, the results
we discuss are valid for all three models of hysteresis.%
  \footnote{In the short--range model, we concentrate on the case where
  we measure only near $R_c$ and $H_c$: we do not integrate over the loop.
  Integrating over the loop yields $\tau$ very close to two, where
  both forms of the scaling should compete.}
See also~\cite{DuriZapp04} in this volume for a more complete discussion 
of our power spectrum theory and the experimental context.

There have been a number of naive arguments for the power law for $E(\omega)$,
stretching back to 
1972~\cite{LienGros72,PerkDahmSeth95,DahmSeth96,SpasBukvMilo+96}, which yield
a power law which is only valid if $\tau>2$:
\begin{equation}
\label{eq:NaiveEnergySpectrum}
E_{\mathrm{naive}}(\omega) \propto \omega^{-(3-\tau)/\sigma\nu z}
	~~~~(\mathrm{only~for~}\tau>2).
\end{equation}
Let $S_{\mathrm{max}}$ be the typical largest avalanche in the system,
say because the system is finite in size.
At root, these derivations fail for $\tau<2$ because 
\begin{equation}
\label{eq:NaiveSIntegral}
\int_0^{S_{\mathrm{max}}} S {\cal D}(S) \,dS 
	= \int_0^{S_{\mathrm{max}}} S^{1-\tau} \,dS 
	= S_{\mathrm{max}}^{2-\tau} 
\end{equation}
diverges. Since there are $S$ spins in each avalanche, it would seem that
equation~\ref{eq:NaiveSIntegral} should represent the number of spins in
the system. This is roughly correct for $\tau>2$, but for $\tau<2$ the
amplitude of the power law $D(S)= D_0 S^{-\tau}$ depends on this cutoff
$D_0 \sim S_{\mathrm{max}}^{-2}$. Continuing a detailed 
analysis~\cite{KuntSeth00}, we get a different scaling form
\begin{equation}
\label{eq:KuntzEnergySpectrum}
E_{\mathrm{naive}}(\omega) \propto \omega^{-1/\sigma\nu z}
	~~~~(\mathrm{for~}\tau<2).
\end{equation}
It is not a surprise that this new prediction fixes a number of discrepancies
between theory and experiments~\cite{KuntSeth00}.

Figure~\ref{fig:PowerScaling} shows the power spectrum $P(\omega)$ 
(equal to $E(\omega)$ divided by the duration of the measurement), together
with four other curves, for both the front--propagation model and the 
infinite--range model~\cite{KuntSeth00}. The top two curves are the
naive power law (equation~\ref{eq:NaiveEnergySpectrum}, obviously not correct)
and the correct law (equation~\ref{eq:KuntzEnergySpectrum}). 
We compare with the relation
between the avalanche size $S$ and the duration $T$ (equation~\ref{eq:TvsS}),
which has the
same combination of critical exponents $S\sim (1/T)^{-1/\sigma\nu z}$. 
We also show that the energy spectrum of individual
large avalanches is proportional to $S$ and has the same power law in 
frequency $\omega$ as that of the entire time series, so we plot
$P(\omega|S) / S$ for a variety of avalanche sizes $S$. 

\vfill\eject

\begin{widetext}

\subsection{Tables of Results}

Here we summarize, from~\cite{PerkDahmSeth99}, our numerical estimates
of the universal critical exponents and various non--universal quantities
used in the scaling collapses above.%
  \footnote{Some of the scaling collapses shown used exponents slightly
  different from those in the tables: see~\cite{PerkDahmSeth99}.}

%\begin{table}
%\narrowtext
\centerline{\begin{tabular}{cr@{$\,\pm\,$}lr@{$\,\pm\,$}lr@{$\,\pm\,$}lc}
measured exponents&
\multicolumn{2}{c}{$3$d} &
\multicolumn{2}{c}{$4$d} &
\multicolumn{2}{c}{$5$d} &
\multicolumn{1}{c}{mean field}  \\ \hline
$1/\nu$ & 0.71 & 0.09 & 1.12 & 0.11 & 1.47 & 0.15 & 2 \\
$\theta$ & 0.015 & 0.015 & 0.32 & 0.06 & 1.03 & 0.10 & 1 \\
$(\tau+\sigma\beta\delta -3)/\sigma \nu$ & -2.90 & 0.16
& -3.20 & 0.24 & -2.95 & 0.13 & -3 \\
$1/\sigma$ & 4.2 & 0.3 & 3.20 & 0.25 & 2.35 & 0.25 & 2 \\
$\tau+\sigma\beta\delta$ & 2.03 & 0.03 & 2.07 & 0.03 & 2.15 & 0.04 & 9/4 \\
$\tau$ & 1.60 & 0.06 & 1.53 & 0.08 & 1.48 & 0.10 & 3/2 \\
$d + \beta/\nu$ & 3.07 & 0.30 & 4.15 & 0.20 & 5.1 & 0.4 & 7 (at
$d_c=6$)\\
$\beta/\nu$ & 0.025 & 0.020 & 0.19 & 0.05 & 0.37 & 0.08 & 1 \\
$\sigma\nu z$ & 0.57 & 0.03 & 0.56 & 0.03 & 0.545 & 0.025  & 1/2 \\
\end{tabular}}
\stepcounter{table}
% \label{table:exponents}
\noindent{Table~\Roman{table}. {\bf Measured universal critical exponents.}
Values for the exponents extracted from scaling collapses in $3$, $4$,
and $5$ dimensions. The mean field values are calculated
analytically~\protect\cite{SethDahmKart+93,DahmSeth93}. $\nu$ is the correlation
length exponent and is found from collapses of avalanche correlations,
number of spanning avalanches, and moments of the avalanche size
distribution data. The exponent $\theta$ is a measure of the number of
spanning avalanches and is obtained from collapses of that data.
$(\tau+\sigma\beta\delta-3)/\sigma\nu$ is obtained from the second
moments of the avalanche size distribution collapses. $1/\sigma$ is
associated with the cutoff in the power law distribution of avalanche
sizes integrated over the field $H$, while $\tau+\sigma\beta\delta$
gives the slope of that distribution. $\tau$ is obtained from the binned
avalanche size distribution collapses~\cite{PerkDahmSeth99}. 
$d+\beta/\nu$ is obtained from
avalanche correlation collapses and $\beta/\nu$ from magnetization
discontinuity collapses. $\sigma\nu z$ is the exponent combination for
the time distribution of avalanche sizes and is extracted from that
data. Error bars are based on variations in the results based on different
approaches to the analysis: statistical fluctuations are typically smaller.
}
%\end{table}
\vspace{0.25cm}

%\begin{table}
\centerline{
\begin{tabular}{cr@{$\,\pm\,$}lr@{$\,\pm\,$}lr@{$\,\pm\,$}lc}
calculated exponents &
\multicolumn{2}{c}{$3$d} &
\multicolumn{2}{c}{$4$d} &
\multicolumn{2}{c}{$5$d} &
\multicolumn{1}{c}{mean field}  \\ \hline
$\sigma \beta \delta$ & 0.43 & 0.07 & 0.54 & 0.08 & 0.67 & 0.11 &
3/4 \\
$\beta\delta$ & 1.81 & 0.32 & 1.73 & 0.29 & 1.57 & 0.31 & 3/2 \\
$\beta$ & 0.035 & 0.028 & 0.169 & 0.048 & 0.252 & 0.060 & 1/2 \\$\sigma\nu$ & 0.34 & 0.05 & 0.28 & 0.04 & 0.29 & 0.04 & 1/4 \\
$\eta = 2 + (\beta-\beta\delta)/\nu$ & 0.73 & 0.28 & 0.25 & 0.38 &
0.06 & 0.51 & 0 \\
\end{tabular}}
\stepcounter{table}
% \label{table:calculated_exp}
\noindent{Table~\Roman{table}. {\bf Calculated universal critical exponents.}
Values for exponents in $3$, $4$, and $5$ dimensions that are
not extracted directly from scaling collapses, but instead are derived
from Table~I and the exponent relations
(see section~\ref{sec:ExponentRelations}). The mean field values are obtained
analytically~\protect\cite{SethDahmKart+93,DahmSeth93}. 
Both $\sigma\beta\delta$ and
$\beta\delta$ could have larger systematic errors than the errors listed
here~\cite{PerkDahmSeth99}. Recent work~\cite{PereVive03} on small
systems, but with sophisticated scaling analysis, suggests a larger value
for $\beta$.}
%\end{table}
\vspace{0.25cm}

%\begin{table}
%\narrowtext
\centerline{\begin{tabular}{cr@{$\,\pm\,$}lr@{$\,\pm\,$}lr@{$\,\pm\,$}lc}
 &
\multicolumn{2}{c}{$3$d} &
\multicolumn{2}{c}{$4$d} &
\multicolumn{2}{c}{$5$d} &
\multicolumn{1}{c}{mean field} \\ \hline$R_c$ & 2.16 & 0.03 & 4.10 & 0.02 & 5.96 & 0.02 & 0.79788456 \\
$H_c$ & 1.435 & 0.004 & 1.265 & 0.007 & 1.175 & 0.004 & 0 \\
$B$ & 0.39 & 0.08 & 0.46 & 0.05 & 0.23 & 0.08 & 0 \\
\end{tabular}}
\stepcounter{table}
% \label{table:RH}
\noindent{Table~\Roman{table}. {\bf Non-universal scaling variables.}
Numerical values for the critical disorders and fields, and the
rotation parameter $B$ (equation~\ref{eq:scaling_2} and subsequent discussion),
in $3$, $4$, and $5$ dimensions extracted from scaling collapses. The
critical disorder is obtained from collapses of the spanning avalanches
and the second moments of the avalanche size distribution. The critical
field is obtained from the binned avalanche size
distribution~\cite{PerkDahmSeth99} and the
magnetization curves. $H_c$ is affected by finite sizes, and systematic
errors could be larger than the ones listed here. The mean field
values are calculated analytically~\protect\cite{SethDahmKart+93,DahmSeth93}.
The rotation $B$ is obtained from the $dM/dH$ collapses. 
}
%\end{table}
\vspace{0.25cm}

\end{widetext}

\section{Exponent relations}
\label{sec:ExponentRelations}

In the following sections we list various exponent relations
for the nonequilibrium RFIM, for which
we give detailed arguments in~\cite{Dahm95,DahmSeth96}.

\subsection{Exponent equalities}

The exponents introduced above are related by the following exponent
equalities:

\begin{equation}
\beta - \beta\delta = (\tau - 2)/\sigma\ {\rm if}\ \tau < 2,
\end{equation}

\begin{equation}
(2 - \eta)\nu = \beta\delta - \beta,
\end{equation}

\begin{equation}
\beta = \frac{\nu}{2} (d - 4 + \bar{\eta}),
\end{equation}

and

\begin{equation}
\delta = (d-2\eta + \bar{\eta})/(d-4+\bar{\eta}).
\end{equation}

(The latter three equations are not independent and are also valid in the
equilibrium random-field Ising model).

\subsection{Two violations of hyperscaling}

In the nonequilibrium RFIM there are two different violations of hyperscaling.

(1) In~\cite{Dahm95,DahmSeth96}, we show that the connectivity 
hyperscaling
relation $1/\sigma = d\nu -\beta$ from percolation is violated in our
system. There is a new exponent $\theta$ defined by

\begin{equation}
1/\sigma = (d-\theta )\nu -\beta
\end{equation}

\noindent with $\theta\nu = 1/2 - \epsilon /6+0(\epsilon^2)$ and
$\theta\nu = 0.021 \pm 0.021$ in three dimensions. $\theta$ is related
to the number of system spanning avalanches observed during a sweep
through the hysteresis loop: see also~\cite{PereVive03}. 

(2) As we discuss in~\cite{DahmSeth96} there is a mapping of the
perturbation theory for our problem to that of the equilibrium
random-field Ising model to all orders in $\epsilon$. From that mapping
we deduce the breakdown of an infamous (``energy")-hyperscaling relation

\begin{equation}
\beta + \beta\delta = (d-\bar{\theta})\nu,
\end{equation}

\noindent with a new exponent $\bar{\theta}$, which has caused much
controversy in the case of the {\it equilibrium} random-field Ising
model.

In~\cite{Dahm95,DahmSeth96}
we discuss the relation of the exponent $\bar{\theta}$
to the energy output of the avalanches. The $\epsilon$ expansion yields
$\bar{\theta} = 2$ to all orders in $\epsilon$. Nonperturbative
corrections are expected to lead to deviations of $\bar{\theta}$ from 2
as the dimension is lowered. The same is true in the case of the
equilibrium RFIM. The numerical result in three dimensions is
% XXX Karin: I changed this? Somehow 1.5 got changed to 1/5, which got me 
% checking it. In your PRB epsilon-expansion paper it's 1.5, not 1/5, but 
% when I do the calculation from the table in Olga's PRB it's 1.69, with 
% a smaller error bar.
%$\bar{\theta} = 1/5 \pm 0.5$. 
$\bar{\theta} = 1.69 \pm 0.28$. 
(In the three-dimensional equilibrium RFIM
\cite{MiddFish02} it is $\bar{\theta}_{eq} = 1.5 \pm 0.4$).

Another strictly perturbative exponent equality, which is also obtained
from the perturbative mapping to the random-field Ising model is given by

\begin{equation}
\bar{\eta} = \eta.
\end{equation}

\noindent It, too, is expected to be violated by nonperturbative
corrections below six dimensions.

\subsection{Exponent inequalities}

In~\cite{Dahm95,DahmSeth96} we give arguments for the following two
exponent-inequalities.%
  \footnote{From the normalization of the avalanche size
  distribution $D(s,r,h)$ it follows that $\tau > 1$.}
% XXX Karin: I moved this expression in brackets into a footnote, which is
% what you had in your PRB. But perhaps it should be a third equation in the
% list?

\begin{equation}
\nu/\beta \delta \geq 2/d,
\end{equation}

\noindent which is formally equivalent to the ``Schwartz-Soffer"
inequality, $\bar{\nu}\leq 2\nu$, first derived for the equilibrium
random-field Ising model, and

\begin{equation}
\nu\geq 2/d,
\end{equation}

\noindent which is a weaker bound than Eq. (3) so long as
$\beta\delta\geq 1$, as appears to be the case both theoretically and
numerically at least for $d\geq 3$.

\section{Finite Sweeprate}

Originally the nonequilibrium RFIM was studied
in the adiabatic limit, where the external field is kept 
constant during each spin flip avalanche and only increased
between subsequent avalanches. In real experiments, however,
the driving field $H$ is typically increased at a finite rate
$\Omega$ such that $H=\Omega t$ where $t$ is time.
This finite rate allows for new avalanches to be triggered by the 
increasing external field while an earlier avalanche is still propagating.
In~\cite{WhitDahm03,TravWhitDahm01} the effect of finite field sweep rates
on the power spectra and avalanche size distributions is
discussed in detail for a large class of systems with crackling noise.
In particular, it is asked how the scaling
behavior of the avalanche size and duration distribution and the 
power spectra of Barkhausen noise or crackling noise in general depend on the 
field sweep rate $\Omega$.
One of the results is an exponent inequality as a criteria for 
the relevance
of adding a small driving rate $\Omega > 0$ to the adiabatic case
$\Omega \rightarrow 0$: If in the adiabatic case
the avalanche duration distribution scales as $D(T)\sim T^{-\alpha}$ with
$\alpha=2$ (such as in the ABBM model or the mean field version of the
RFIM), then at (small enough) finite field sweep rate the corresponding
noise ``pulse'' duration distribution is expected to scale as
$D(T,\Omega) \sim T^{2 -a(H)C \Omega}$ where $a(H)$ and $C$ are nonuniversal
constants~\cite{DuriBertMagn95,BertDuriMagn94,Duri97,DuriCizeZapp+98}. 

If, however, in the adiabatic limit the exponent $\alpha$ 
is either greater or smaller than $2$, then at (small) finite field sweep rate 
the pulse duration distribution exponent remains the same as in the
adiabatic limit. Note that in the case $\alpha <2$ especially,
$\Omega$
has to be particularly small in order to be able to still see distinct pulses
even at finite field sweep rate, since at higher sweep rate the system 
will  very quickly develop runaway events. In~\cite{WhitDahm03,TravWhitDahm01}
quantitative criteria for the meaning of ``small sweeprate'' are given. Also,
the zero temperature nonequilibrium RFIM 
and recent variants are used to numerically test the analytic results, 
which are expected to be 
applicable to a much larger class of systems 
with crackling noise (the exact conditions are discussed 
in~\cite{WhitDahm03,TravWhitDahm01}).
It is found that the results agree well with both simulations and
recent experiments on Barkhausen noise in various soft magnetic 
materials with and without applied stress (with $\alpha < 2$ and $\alpha = 2 $ 
respectively~\cite{DuriZapp00}).
A brief review of other, related studies of the effects of finite field 
sweep rate is also given in~\cite{WhitDahm03}.

\section{Subloops and History Induced Critical Behavior}

One of the characteristic features of magnetic hysteresis are the subloops,
seen as the external field is changed up and down an amount insufficient
to saturate the magnetization (figure~\ref{fig:HysteresisLoops}). 

\subsection{Return--Point Memory}

Much attention has been paid to modeling these subloops with the Preisach
model~\cite{Mayergoyz}
Preisach models are quite different in spirit to ours: they represent
the system as a large number of uncoupled hysteretic two--state domains,
and fit the distribution to the observed behavior. They are able to model
a hysteretic system if it possesses 
{\em return--point memory}~\cite{BertBook}, also
known as {\em wiping out}~\cite{Maye86}. 

Return--point memory states that the magnetization after a subloop rejoins
a larger loop (perhaps itself a subloop) equals the magnetization at which
the subloop left the outer loop. That is, if the subloop represents an
excursion downward from a local maximum external field $H$, then when $H$
returns to its previous maximum the magnetization $M$ returns to its
previous value -- remembering its previous magnetization on returning, and 
wiping out all effects of the excursion. Return--point memory states that
the subloops in figure~\ref{fig:HysteresisLoops} should close perfectly,
without a gap and without crossing themselves (as observed).

Indeed, our interacting, three-dimensional, disordered models exhibit
the return--point memory in an even stronger form. Upon rejoining the
larger loop, the entire state of the system is microscopically identical
to the state it had when it left the larger loop (a microscopic return--point
memory, as opposed to a macroscopic memory constraining only the net
magnetization). In~\cite{SethDahmKart+93}, we showed in great generality
that return point memory should be true of any system with
\begin{enumerate}
\item
{\bf A partial ordering of states.}
Here, one microstate $A$ is ``ahead'' of another
$B$, $A>B$, if every spin in $A$ is greater than or equal to the 
corresponding spin in $B$.
\item
{\bf No passing.} 
That is, the dynamics preserves the partial ordering.
\item
{\bf Adiabatic.}
The external field is raised and lowered slowly enough that the system
does not lag behind.
\end{enumerate}

Return--point memory can be quite remarkable: in our model (and in 
some experimental systems) repeating a subloop plays back precisely the
same Barkhausen noise. Indeed, the return--point memory
can be used as a way of retrieving analogue magnetic memories that is
(at least theoretically) significantly superior to measuring the
remanent magnetization~\cite{PerkSeth97}.
On the other hand, the absence of exact (microscopic) return point
memory has also been observed~\cite{DellTorrVajd95,Guil70,ZappDuri01}, and
there are various interesting experiments testing for the reproducibility 
of magnetic avalanches~\cite{UrbaMadiMarkRPM95,PettWeisDuri97,PettWeisOBri96}
and for microscopic return point memory and complementary 
point memory in magnets at various disorders~\cite{PierBuecSore+04}.%
  \footnote{The complementary point memory relates the magnetic domains
  at one point on the major hysteresis loop to the domains at the 
  complementary point on the major loop during the same and during subsequent
  cycles.}
In~\cite{PierBuecSore+04} the authors experimentally study the 
influence of disorder on major loop return point memory and complementary 
return point memory in Co/Pt samples with varying interfacial roughness
and find with increasing disorder the onset and saturation of both
return point memory and complementary point memory.

Disorder dependence of return point memory versus reptation ({\it i.e.}
is gradual subloop closure~\cite{Neel58,Neel59}) has 
recently been reported for the zero temperature random coercivity model with 
antiferromagnetic-like interactions~\cite{HovoFried03}.

A related memory effect, recently reported for the hysteresis of 
spin glasses, is the reversal field memory effect~\cite{KatzPazmPike+02}.
Katzgraber {\it et al.} show that this memory effect emerges in the
nonequilibrium Edwards Anderson spin glass (EASG) when the magnetic field is
first decreased from its saturation value and then increased again from some
reversal field $H_R$. The authors find that EASG exhibits a singularity
at the negative of the reversal field, $-H_R$, in the form of a kink
in the reversal of the magnetization of the reversal curve. They show 
that this memory effect 
is due to a local spin-reversal symmetry of the Hamiltonian.
This symmetry and thus the reversal field memory effect is present 
in spin glasses where the disorder 
is due to random couplings between spins. In general it is not expected 
in systems 
with random magnetic fields such as the RFIM, since there the disorder breaks 
local spin reversal symmetry.

As an extension to memory effects for a driving field that varies 
in two dimensions, 
a vector form of wipe-out memory for situations when the magnetic field 
varies in both direction and magnitude, has been suggested
based on the two dimensional vector Preisach model in~\cite{Frie99},
and has been experimentally tested in~\cite{FrieChaHuan+00}.

\subsection{Critical Behavior in Subloops}

\begin{figure}[thb]
  \begin{center}
    \epsfxsize=8cm
    \epsffile{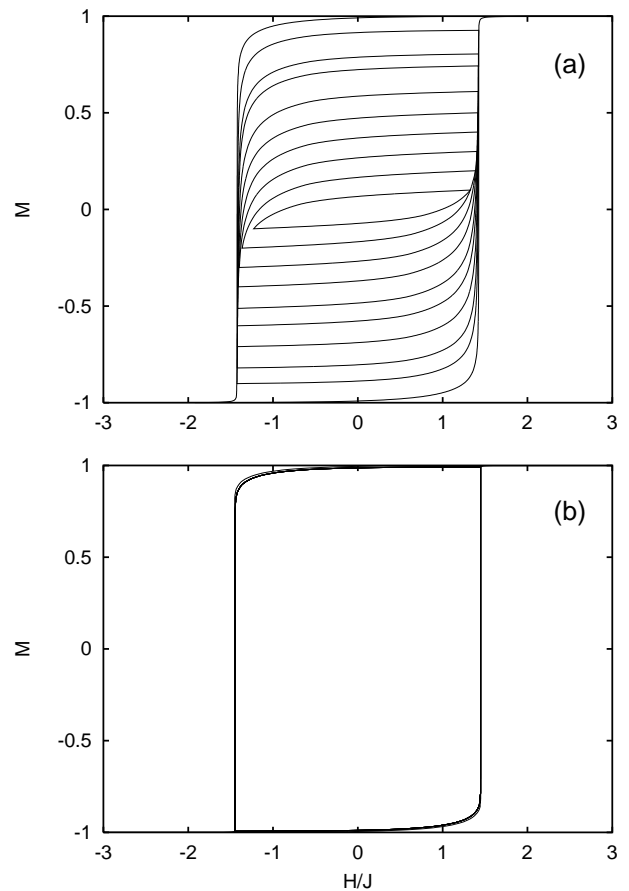}
  \end{center}
\caption{{\bf Saturation hysteresis loops with concentric inner subloops.}
Top: $R=2.225J > R_c$; bottom: $R=2.1<R_c$.~\cite{CarpDahm01}
} 
\label{fig:Subloops}
\end{figure}

The disorder induced critical scaling for hysteresis loops in the RFIM
that we discussed so far all referred to the saturation hysteresis
loop. However, often it is experimentally
impractical to take magnets to their saturation point due to the large
magnetic fields required, so the behavior of subloops
(figure~\ref{fig:Subloops},~\cite{CarpDahm01}) is of great
interest to experiments and applications. The RFIM can be used
to model subloops, and in one dimension~\cite{Shuk00} and on a Bethe
lattice~\cite{DharShukSeth97,ColaGabrZapp02} they have been computed 
exactly~\cite{Shuk01}. The magnetization
curves of subloops have been collapsed near the demagnetized state
using Rayleigh's law~\cite{DuriZapp01,ZappMagnDuri02}. 
Interestingly, in the RFIM not only the saturation hysteresis loops but also 
subloops reflect the critical point described above, 
and at the critical disorder one finds
{\it history-induced} critical scaling. Concentric inner subloops resemble
rescaled saturation loops at effectively higher (possibly correlated)
disorder. Avalanche size distributions for the inner subloops, too, look
similar to rescaled avalanche size distributions of saturation loops
at higher disorder. We have studied universal properties of this history 
induced critical
behavior using numerical simulations of systems with more than 
100,000,000 spins and Widom scaling methods and extracted universal 
scaling exponents and scaling functions. These predictions are expected to 
apply to experiments on hysteresis and Barkhausen noise in hard magnetic 
materials. To model soft magnets
on the other hand, we added demagnetizing fields to the model and obtained
single domain wall propagation, as described in~\cite{KuntSeth00}. 
As a result the
system was always critical, even for the subloops. The results of this
study are discussed in~\cite{CarpDahm01,CarpDahm04}.
A special, related kind of subloops -- so called ``first order reversal 
curves'' (FORC) -- have proven very useful in distinguishing between different
magnetic materials, see~\cite{KatzGaryZima04}.

\subsection{Demagnetization Curves}

The RFIM also predicts disorder induced critical behavior in the 
demagnetization curve, or initial magnetization curve and the associated
Barkhausen noise.  Ferromagnetic 
materials with a remanent magnetization can be demagnetized by applying an
oscillating magnetic field with amplitude slowly decreasing from a
large initial value to zero. Sometimes the final state is also termed
AC-demagnetized state. The oscillating external field with decreasing
amplitude takes the system through concentric subloops.
The line connecting the tips of the subloops is
known as the {\it normal or initial magnetization curve}, or {\it
demagnetization curve}. It is to be distinguished from the {\it
virgin curve} which is obtained by thermal 
demagnetization~\cite{BertBook,FronVive02,FronVive00}.
One can use the RFIM to simulate the
demagnetization curve for different amounts of disorder, especially
near the critical disorder mentioned above for the saturation loop. 
It turns out 
that this curve reflects much of the scaling behavior found for the saturation
hysteresis loop. This may be surprising, since the meta-stable states
encountered in the demagnetization curve are completely different from
those of the saturation loop. In~\cite{CarpDahm03}
the model is briefly described
and the necessary exponent relations are derived, which 
show that the exponents for the demagnetization curve
can actually be derived from those of the
subloops. Predictions for the exponents from simulations of up to 
one million spins are presented.
The demagnetization curve in the presence of long range demagnetizing fields 
is also discussed, with resulting predictions for experiments on 
demagnetization curves and associated Barkhausen noise in soft magnetic 
materials. 
For a discussions of the Rayleigh law and scaling of the demagnetization
curve at small magnetic fields, see~\cite{ZappMagnDuri02,ColaGabrZapp02}.
Zarand, Pazmandi, Pal, and Zimanyi recently proposed a new optimization
method to search for ground states, which is based on the 
demagnetization procedure~\cite{ZaraPazmPal+02}.

\section{Real Experiments}

% XXX Karin #2: Either mention more recent magnetism experiments, or 
% explain the more recent experiments (or experiments on soft materials) 
% should refer to Durin.
% XXX - hard and soft magnets ....

\subsection{Magnetic hysteresis loops}

There are a few recent experiments on hysteresis and crackling noise
where disorder is tuned and evidence of a phase transition related to 
that of the nonequilibrium RFIM is reported:
In~\cite{BergPangHops00}, Berger \etal describe
a transition from smooth magnetic hysteresis loops 
to those with a jump in the magnetization in
ultrathin epitaxial Gd-films with in-plane magnetization upon 
annealing the sample
at successively higher annealing temperatures. The authors show that
annealing at higher temperature increasingly reduces the disorder in 
the crystal structure of the films. In
\cite{BergInomJian+00,BergInomJian+01,BergPechComp+01} Berger \etal report a similar transition
for thin Co/CoO films.
In this system the antiferromagnetic CoO layer allows a reversible tuning 
of the magnetic disorder felt by the ferromagnetic Co layer, by simple 
temperature variation. The authors
are able to extract scaling collapses for the magnetization hysteresis loops
at different disorders, and find the values of the associated scaling 
exponents to be $\beta = 0.022 \pm 0.006$ and $\beta\delta = 0.30 \pm 0.03$. 
Note that this magnetic system is effectively two dimensional.

Incidentally, almost the same exponent values ($\beta = 0.03 \pm 0.01$
and $\beta \delta= 0.4 \pm 0.1$) are reported by Marcos \etal 
in~\cite{MarcViveMano+03} for the magnetic hysteresis loops of Cu-Al-Mn
alloys with different Mn content at low temperatures. The loops are 
smooth above a certain temperature, but exhibit a jump in the 
magnetization below that temperature. The authors perform a scaling 
analysis near this critical point with temperature as the effective 
``disorder''  tuning parameter. They point out that, contrary to the 
thin Co/CoO bilayers in Berger \etal~\cite{BergInomJian+00},
this system is effectively three dimensional.
 
For these three systems it would be very interesting to also see
a scaling analysis of the associated Barkhausen noise, since 
from simulations we know that the exponents extracted from
avalanche size distributions and noise power spectra are often more
reliable than those obtained just from magnetization curves 
alone~\cite{DahmSeth96}. For these three systems, so far, only the 
scaling of the magnetization curves has been reported.

\subsection{Disorder effects on Barkhausen Noise}

Disorder effects on crackling noise (rather than hysteresis loops) 
has been seen in other systems. Barkhausen noise is commonly used for 
nondestructive testing -- in engineering contexts we find careful studies 
of the effect of the amount of carbon content in steel on Barkhausen noise.
One might speculate that varying the carbon content could
be similar to varying the effective disorder in the system. It would be
interesting to see a systematic scaling analysis for the available
experimental data.
In an early study~\cite{LienGros72} Lieneweg and Grosse-Nobis found
that the distribution of Barkhausen pulse areas integrated over the 
hysteresis loop of an 81\% Ni-Fe wire was well described by a power 
law (with exponent $\tau+\sigma\beta\delta$ ranging from $1.73$ to 
$2.1$) up to a certain cutoff size. Annealing the sample at various
annealing temperatures they found that the cutoff appeared to be smaller
at higher annealing temperatures. It would be interesting to see 
whether the cutoff grows to takes a (system-size dependent) maximum value at a critical
annealing temperature $T_c^{ann}$ and decreases again at higher and lower
annealing temperatures, as one would expect if the RFIM applies with
temperature as the effective disorder tuning parameter in this system.
(In leading approximation, near $T_c^{ann}$ they should be linearly related to 
each other.)
Near $T_c^{ann}$ the Barkhausen pulse area distribution should then 
be described by a scaling form that would allow a scaling collapse
of all distributions onto one single curve for appropriate rescaling
of the axes. Again, potentially universal critical
exponents could be extracted from such a collapse. They would be predicted 
by the RFIM avalanche critical exponents if that model is in the same 
universality class.

There are other experiments which revealed power law decays
for Barkhausen pulse size distributions in various samples, as
reviewed for example in~\cite{SethDahmMyer01} and in the article by Durin
and Zapperi in this book~\cite{DuriZapp04}. Many of these are measurements
in soft magnetic materials where long range interactions are important
and the main mechanism for Barkhausen noise is domain wall propagation
rather than new domain nucleation. Disorder
has usually not been varied in these experiments. Models predict that
these systems naturally tend to operate close
to the underlying critical domain wall depinning transition and thus 
produce power law pulse size
distributions regardless of the underlying disorder, 
(at least as long as it 
is less than a critical value above which the domain wall may have
overhangs
\cite{JiRobb92,UrbaMadiMark95,Nara96,ZappCizeDuri+98,DuriZapp00,DuriZapp01}).
The disorder induced critical point of the nonequilibrium RFIM
is expected to be relevant for hard magnetic materials instead,
where either the disorder is very strong compared to the effects
of long range dipolar interactions, or the geometry of the 
sample and experiment is chosen such that the effects of long range
dipolar fields are small (as for thin films with in-plane applied 
magnetic field, or for thin wires with parallel applied field).

\subsection{Imaging magnetic avalanches and states}

\begin{figure}[thb]
  \begin{center}
    \epsfxsize=8cm
    \epsffile{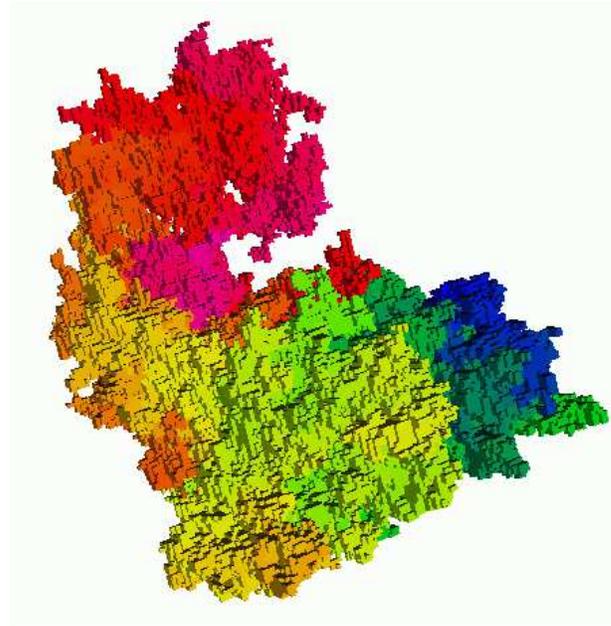}
  \end{center}
\caption{{\bf Fractal spatial structure of an avalanche.} This moderate--sized
avalanche contains 282,785 domains~\cite{SethDahmMyer01}.
} 
\label{fig:BigAvalanche}
\end{figure}

The avalanche structures predicted by the theoretical models are visually
interesting (figure~\ref{fig:BigAvalanche}); in our model they have
fractal dimensions $1/\sigma\nu$ close to, but probably less than three.
Exciting experiments probing these spatial structures are coming on line.

Recently interesting experiments imaging avalanches in magnetic films 
have been reported. In~\cite{DKimChoeShin03} Kim, Choe, and Shin report
full-field magneto-optical observations (via Kerr effect) of Barkhausen 
avalanches in Co polycrystalline thin films at criticality, i.e. their 
imaged avalanches are due to single domain wall propagation rather than
domain nucleation. In this 
experiment many avalanches extend far beyond the imaged region, so that
their measured exponent appears to be an exponent combination
of the exponents computed in the models.
In addition to their results it'd be interesting to also see time
resolved images of entire avalanches and direct measurements of the
roughness exponent of the propagating domain wall, to compare with 
predictions from models for magnetic domain wall propagation.
The magnetization reversal dynamics of epitaxial Fe films grown on 
GaAs(001) and the field dependent evolution of domain structure at 
various field has also been studied using the magneto-optic Kerr effect,
see for example~\cite{LeeChoiXuBland99}.
In~\cite{SchwLieb+04} Schwarz {\it et al.} report on visualization
of the Barkhausen effect by magnetic force microscopy for a
granular thin film with perpendicular anisotropy. The authors are able
to localize individual Barkhausen volumes and distinguish them as either
newly nucleated or grown by domain wall propagation. They find a Gaussian
size distribution of nucleated Barkhausen volumes indicating an uncorrelated
random process, and a power law distribution for grown Barkhausen volumes, 
reflecting critical behavior during domain wall motion.
Another very interesting experimental approach 
is x-ray speckle metrology~\cite{PierBuecSore+04}. In a recent preprint
Pierce {\it et al} use this method to study correlations
between microstates with different histories and test for microscopic
return point memory of subloops in magnetic systems.

\subsection{Random Bonds and Random Anisotropy}

Renormalization group treatments, symmetry arguments~\cite{DahmSeth96},
and simulations~\cite{VivePlan94,ViveGoicOrti+95} show, that the hysteretic, 
zero temperature
random bond Ising model with ferromagnetic mean bond strength is expected 
to be in the same universality class as the nonequilibrium RFIM. Vives
{\it et al.} also point out that their numerical scaling exponents in
two dimensions of the two models also have overlapping error bars
with the corresponding 
exponents of Blume-Emery Griffiths models with disorder, in which 
the spins take three values $\pm 1,0$~\cite{ViveGoicOrti+95}.
An analysis of the short time dynamics of first-order
phase transitions in a disordered system, using the random bond 
Ising model is given in~\cite{Zhen02}. 
The diluted ferromagnetic random Ising model~\cite{Tadic96} has both
nonuniform bond strengths that reflect dilution due to nonmagnetic 
impurities, and random magnetic fields. 
Symmetry arguments suggest~\cite{DahmSeth96} that a model with uncorrelated 
random (ferromagnetic) bonds and random fields should again be
in the same universality class as the RFIM, since it has no new symmetries
compared with the RFIM.
Similarly, daSilveira and Kardar developed a $O(N)$ vector spin model with 
random fields and showed through a renormalization group
calculation that, in general, it can displays the same continuous  
phase transition in the hysteresis loop as the RFIM~\cite{daSiKard99},
{\it i.e.} the two critical points are in the same 
universality class.%
  \footnote{They also find a new universality class, see 
  section~\ref{sec:Unsolved}.}
The reason is that the applied external field breaks 
any added rotational symmetry~\cite{DahmSeth96} so that the behavior
on long length scales is determined by the same the same symmetries in both
models. Similarly the random anisotropy model of~\cite{VivePlane01},
was shown numerically to be in the 
same universality class as the RFIM: this is again as expected, since
the external applied field
not only breaks rotational symmetry but also time-reversal symmetry.
Perhaps somewhat surprisingly, a transition from hysteresis loops
with a jump  to those without, like the one of the RFIM, is also seen in a 
plastic depinning model for charge density waves~\cite{MarcDahm02}. 
If in the random field Ising model
a certain fraction of the bonds between neighboring spins is enhanced 
to a very large value, one obtains a simple model for exchange bias
\cite{IllaVivePlane02}.

Hysteresis loops observed for the Ising glass with zero ferromagnetic 
mean bond strength are smooth and continuous~\cite{PazmZara+99,BanePuri+01}
-- qualitatively this agrees with experimental hysteresis loops seen 
for example in AuFe systems~\cite{BorgKitc73}.
For Ising glasses with strong ferromagnetic mean bond strength (compared
to the width of the distribution of bond strengths, which is the disorder),
a jump in the magnetization curve appears, just as in the low disorder
phase of the RFIM. The hysteresis loops resemble those seen for more
concentrated AuFe systems~\cite{BorgKitc73} as well as CuMn and AgMn
systems~\cite{MonoPrej+79,PrejJoli+80}.
A brief review of these and other glassy spin systems
such as Ising spin glasses with long-range interactions and 
Heisenberg spin glasses with long-range interactions in both theory
and experiment is given for example in~\cite{BanePuri+01}.

An interesting hysteresis model with rather different properties than
ours based on an infinite--range spin glass (Sherrington-Kirkpatrick model)
is reported to show self-organized criticality; see~\cite{PazmZara+99} 
for details.

\subsection{Other dynamics at high sweep rates}

Much interesting work has been done in both theory and experiment on 
dynamic hysteresis loops at finite field sweep rate in the presence
of temperature. At field sweep rates higher than a 
critical value that depends on sweep amplitude and temperature, 
the system transitions to a symmetry broken phase with 
a flatter hysteresis loop, centered around a nonzero average 
magnetization. Similar dynamic transitions
also are seen for pulsed and stochastically varying fields. 
Chakrabarti and Acharyya published a review article on both
models and experiments in this regime~\cite{ChakAcha99}.
Some of the addressed questions deal with scaling of the 
hysteresis loop area with  field sweep frequency, amplitude, and 
temperature. While in the initial approaches to these issues
quenched randomness in the system was neglected,
there are recent model approaches and experiments 
studying dynamic hysteresis of a moving domain wall in the presence
of quenched disorder and temperature at finite field sweep 
rate~\cite{NattPokr04,NattPokr+01}.

\subsection{Nonmagnetic noisy hysteretic systems}

In this section we briefly discuss a broad range of related work on
avalanches and hysteresis in non-magnetic systems. We start with a general
overview, and then focus on a few particular systems.

\subsubsection{Systems with avalanches}

Magnets crackle (Barkhausen noise) when the external field is ramped 
upward slowly. Many systems exhibit similar crackling noise: when pushed
slowly they respond with discrete events of a broad range of sizes.%
  \footnote{This subsection follow closely the presentation in 
  reference~\cite{SethDahmMyer01}, with some added references.}
The earth responds~\cite{GutenRich} with violent and intermittent 
earthquakes as two tectonic plates rub past one another. A piece of 
paper~\cite{HoulSeth96}
(or a candy wrapper at the movies~\cite{KramLobk96,Glanz00}) emits
intermittent, sharp noises as it is slowly crumpled or
rumpled. Just like the magnetization avalanches responsible for
Barkhausen noise in magnets, these individual events span many orders of
magnitude in size. Indeed, the distribution of sizes follows a
power law with no characteristic size scale. 

In the past decades, scientists in a variety of fields have been making
rapid progress in developing
models and theories for understanding this sort of
scale-invariant behavior in driven, nonlinear, dynamical
systems. Interest in these sorts of phenomena goes back several
decades. The work of Gutenberg and Richter~\cite{GutenRich} in the 1940's
and 1950's established the well-known frequency-magnitude
relationship for earthquakes that bears their names. 
A variety of many-degree-of-freedom dynamical models 
\cite{BurrKnop67,RiceRuina83,CarlLang89,BakTang89,ChenBakObuk91,OlamFedeChri92,MiltSornVane93,CarlLangShaw94,MyerShawLang96,ShawRice00,BenZRice95,FishDahmRama+97}, with and without disorder, have been introduced in the years
since to investigate the nature of slip complexity in
earthquakes. More recent impetus for work in this field came
from the study of the depinning transition in sliding
charge-density wave (CDW) conductors in the 1980's and early
1990's~\cite{Fish83,Fish85,Litt86,NaraFish92,MiddFish93,MyerSeth93,Thor96}.
Interpretation of the CDW depinning transition as
a dynamic critical phenomenon sprung from Fisher's early
work~\cite{Fish83,Fish85}, and several theoretical and numerical studies
followed. This activity culminated in the RG solution by Narayan
and Fisher~\cite{NaraFish92} and the numerical studies by 
Middleton~\cite{MiddFish93} and
Myers~\cite{MyerSeth93} which combined to provide a clear picture of depinning
in CDWs and open the doors to the study of other disordered,
nonequilibrium systems. 

Bak, Tang, and Wiesenfeld inspired much
of the succeeding work on crackling noise~\cite{BakTangWies87,BakTangWies88}. 
They introduced
the connection between dynamical critical phenomena and
crackling noise, and they emphasized how systems may naturally
end up at the critical point through a process of self-organized
criticality. (Their original model was that of avalanches in
growing sandpiles. Sand has long been used as an example of
crackling noise~\cite{deGennes,Feynman}. However, it turns out that 
real sandpiles
don't crackle at the longest scales~\cite{JaegLiuNage89,Nage92}). 

Researchers have
studied many systems that crackle. Simple models have been
developed to study bubbles rearranging in foams as they are
sheared~\cite{TewaSchiDuri+99}, biological extinctions~\cite{SoleManr96},%
  \footnote{These models are controversial~\cite{Newm96,NewmPalm99}:
  they ignore catastrophic external events like asteroids.}
fluids invading porous materials and other problems involving invading
fronts~\cite{CiepRobb88,KoilRobb00,NattStep+92,NaraFish93,LescNattStep+97,RoteHuchLube+99},%
  \footnote{The random--field model we describe was 
  invented in this context~\cite{CiepRobb88,KoilRobb00}, as described in 
  section~\ref{sec:Models}.} 
the dynamics of superconductors~\cite{FielWitt+95,ErtaKard94,ErtaKard96},
and superfluids~\cite{LillWootHall96,GuyeMcCa96}, sound emitted
during martensitic phase transitions~\cite{OrtiRafoCarri+95}, fluctuations 
in the stock market~\cite{Bouch00,BakPaczShub97,IoriJafa01}, solar flares
\cite{LuHamiMcTi+93}, cascading failures in power 
grids~\cite{CarreNewmDobs+00,SachCarrLync00,Watt02}, failures in systems
designed for optimal performance~\cite{CarlDoyl99,CarlDoyl00,Newm00},
group decision making~\cite{Gala97}, crackling noise in mammalian lungs
\cite{AlenBuldMaju+03}, and fracture in disordered 
materials~\cite{PetrPapaVesp+94,GarcGuarBell+97,CurtSche91,HerrRoux90,ChakBeng97,ZappRayStan+97}.
These models are driven systems
with many degrees of freedom, which respond to the driving in a
series of discrete avalanches spanning a broad range of scales
what we are calling crackling noise.  

There has been healthy skepticism by some established professionals in these
fields to the sometimes grandiose claims by newcomers
proselytizing for an overarching paradigm. But often confusion
arises because of the unusual kind of predictions the new
methods provide. If our models apply at all to a physical
system, they should be able to predict all behavior on long
length and time scales, independent of many microscopic details
f the real world. This predictive capacity comes, however, at a
price: our models typically don't make clear predictions of how
the real-world microscopic parameters affect the
long-length-scale behavior.  In this paper and in 
\cite{SethDahmMyer01} we have provided
overview of the renormalization-group 
\cite{Kada66,Wils79,PfeuToul77,Yeomans92,Fish98} that
many researchers use to understand crackling noise and which
is perhaps the most impressive use of abstraction in science. 

We now turn to some non-magnetic systems that have been particularly 
well studied.

\subsubsection{Martensites}

Hysteresis with crackling noise and in certain cases even
microscopic return point memory has long been seen in ferroelastic materials,
such as shape memory alloys or martensites~\cite{BruceCowley,PlaneMano01}.
A martensitic transformation is a diffusionless first order phase 
transformation where the lattice distortion is mainly described
by a homogeneous shear. Many metals and alloys with a bcc structure 
will upon cooling (or under strain) undergo this transition to a low
temperature close packed structure. In athermal martensites, such as Cu-Zn-Al,
thermal fluctuations are not relevant -- temperature 
acts as an external driving field. Similar to ferromagnets ramping
magnetic field, the martensitic transition takes place as a sequence of
avalanches as the temperature is swept. The analogue of Barkhausen
noise is actual noise: acoustic emission due to elastic waves in the ultrasonic
range, generated by propagating domain walls during the avalanches.

Experimental studies of the acoustic emission generated during a 
thermally induced martensitic transformation of a Cu-Zn-Al single crystal
revealed an apparent absence of characteristic scales in the distribution
of avalanches~\cite{ViveOrti+94}. After an initial number of hysteresis
cycles (ramping temperature) the system reaches a final attractor
yielding two decades of power law scaling in the time distribution of
acoustic emission pulse durations. The authors argue that the cutoff at
large times is probably due to experimental limitations (amplifier
cutoff), and interpret the observed scaling behavior in terms of
self-organized criticality. Of course, the system could have a tunable
parameter that lies close to the critical point (two decades of 
power--law scaling arises in a range 50\% away from our critical point).

In a more recent study, P\'erez-Reche, Stipcich, Vives, Ma\~nosa, Planes,
and Morin~\cite{PereStip+04} 
study the evolution of the kinetic features of the martensitic
transition in a Cu-Al-Mn single crystal under thermal cycling. The authors
use several experimental techniques, including optical microscopy,
calorimetry, and acoustic emission, to perform an analysis at multiple
scales. Focussing especially on avalanches, they find that there are
significant differences between the kinetics at large 
and small scales during the initial cycles. Upon repeated temperature 
cycles, however, the system evolves from displaying a supercritical
avalanche size distribution to a critical power-law distribution.
In the language of the RFIM this could be described as the effective disorder
being increased by the temperature cycles from $R < R_c$ to $R=R_c$.
(The critical exponents, and thus the universality class, is different 
from our RFIM).

Both of these systems demand repeated cycling to produce critical behavior.
This is quite different from systems with return--point memory, where repeated
cycles don't change the behavior. A recently introduced, simple
two dimensional phenomenological model  
for the martensitic transformation is shown to mimic near full reversal of 
morphology under thermal cycling in~\cite{SreeAnan03,AhluAnan01}.

A crossover from the hysteresis loops with a jump or ``burst" to smooth
hysteresis loops (ramping temperature) has been recorded in~\cite{OlsoOwen92},
as the grain size of macroscopic, polycrystalline
specimen of an Fe-Ni-C alloy is reduced. The crossover is explained in terms of
finite size effects in defect induced nucleation within each grain (due
to extended defects like dislocation tangles, however, rather than random
point disorder). It seems clear that one should look for critical
fluctuations near the crossover that would promote an interpretation of
the grain size as the analogue of the disorder parameter $R$ in our
model.

\subsubsection{Liquid Helium in Nuclepore}

In~\cite{LillFinl+93} Lilly, Finley, and Hallock show a hysteresis loop 
for the capillary condensation of
superfluid helium in the porous material Nuclepore. Nuclepore is a
polycarbonate sheet perforated by a high density of nearly cylindrical
holes with a nominal diameter of about 2000$\AA$. The pores intersect.
The most probable number of intersections per pore is about 5. Most
pores participate in a macroscopic interconnected network, (i.e. they are
apparently internally connected beyond the percolation threshold). In
the experiment the authors measure the amount of helium absorbed and
capillary condensed in the Nuclepore sample as a function of the number
of helium atoms present (chemical potential) at fixed temperature $T =
1.52$ K.

Pore intersections result in weak interactions among the pores which lead
to avalanches that have been observed upon slow withdrawal of helium from
the cell. The upper branch of the hysteresis loop 
(upon withdrawal of He from the sample) actually has
microscopic jumps. The largest
avalanche observed involves 2.6 10$^7$ pores, which is about 2.6\% of the
system. The distribution of avalanche sizes is described by
power laws. 

This system has clear analogies to our model. The coupling between the
pores is certainly ``ferromagnetic": If a pore is emptied it becomes
more likely for an intersecting pore to be emptied also. Different pore
radii and different cross-sections of the pore intersections introduce a
certain randomness which might be used to tune the system to a critical
point.
In~\cite{LillHall01} the avalanches are studied for different
characteristic pore sizes and the revelance of various theoretical models
which one might apply to the system is discussed.
In~\cite{DetcKier+04} a unifying theoretical approach to 
hysteresis phenomena
associated with capillary condensation of gases in disodered mesoporous solids
is given, and the connection with the zero-temperature theoretical studies
of magnetization-reversal hysteresis of driven interfaces in disordered 
media is discussed (see also~\cite{DetcKier+03}).

\subsubsection{Superconductors}

\paragraph{Superconducting vortex avalanches}

In an interesting experiment on the dynamics of superconducting vortices 
in the Bean state in $Nb_{47\%}Ti_{53\%}$, Field, Witt and Nori recorded
vortex avalanches as the system was driven to the threshold of
instability by the slow ramping of the external magnetic 
field~\cite{FielWitt+95}. 
The individual avalanches were ranging in size from 50 vortices up to about
$10^7$ vortices, which is about $10^{-6}$ of the total number of vortices
in the system. The distribution of avalanches sizes was recorded for the events
occurring in three 450 G wide windows of the external magnetic field
range, centered at 2.25 kG, 5.33 kG, and 7.55 kG.
At all three values for the field $H$ there seems to be a power law up to
some cutoff size, which varies with $H$.
The authors suggest that the self-organized criticality is the
underlying mechanism, despite the 
fact that their cutoff size changes as a
function of $H$.

\paragraph{Ultra-thin granular superconducting films in a parallel field}

Wu and Adams discovered avalanches in film resistance
corresponding to the collapse of macroscopic superconducting regions
involving up to $10^9$ grains near the parallel critical field in
ultra-thin granular superconducting Al-films~\cite{WuAdams95}. The corresponding
distribution of avalanche sizes displays four decades of power law scaling,
(which in our simulations would correspond to a value of disorder about
10-15\% away from the critical point). Fewer decades are seen at lower
external magnetic fields --- the field may act as a tunable parameter for
a critical point. The authors point out the relation to our model and
the zero temperature random bond Ising model with the same dynamics. The
scaling exponent they extract from the power law decay is similar to the
prediction from simulations of the two dimensional random bond Ising
model.

\subsubsection{Prewetting on a disordered substrate}

Blossey, Kinoshita, and Dupont-Roc have studied the effect of substrate
randomness on the dynamics of a prewetting transition 
\cite{BlosKino+98}. In their theoretical paper they first review
several experiments and then suggest a mapping of 
the hysteretic prewetting system to a two dimensional nonequilibrium
RFIM in an external field $H$, by identifying the Ising spins $\pm 1$
with a wet or nonwet patch of the surface. They find that the covering
of the substrate appears to be a critical transition. The growth exponents
that they extract from their simulations are reported to be compatible
with those of percolation, see also~\cite{SabhDharShuk02}.

\section{Unsolved problems}
\label{sec:Unsolved}

There are many open and challenging questions in this field. 
\begin{enumerate}

\item {\bf Two Dimensions.}
Except for the experiments in magnets and martensites most of these
experiments are in effectively two dimensional systems. $D=2$
appears to be the lower critical dimension of the transition
in the RFIM~\cite{Kunt01,PerkDahmSeth95}. Extracting reliable scaling
behavior in two dimensions remains a challenge, and we have focused here
on three dimensions where our understanding is reasonably complete.

\item {\bf Avalanche average pulse shapes.}
The average pulse shapes predicted by the various theories
(figure~\ref{fig:AmitTheoryPulseShapeCollapse}) do not appear close to
those measured in experiments, which themselves disagree with one another
(figure~\ref{fig:AmitExptPulseShapeCollapse}). This is one of the few
experiments where the scaling function has been extracted from the data.
Are we mistaken about the universality of these functions? Does this
reflect measurement artifacts, or perhaps unrecognized important control
parameters? Are there
analytical corrections to scaling (appendix~\ref{app:ScalingForms}) that
we are missing that skew the curves? Or are there many dynamical universality
classes? Recent experiments~\cite{DurinTalk} suggest that the skewness 
of the smaller avalanches can be affected by external stress.
Signatures of time asymmetries have also been seen in high order spectra
studies of Barkhausen Noise~\cite{OBrieWeiss94,PettWeissDuri98}.

\item {\bf Correlated disorder and long--range fields.}
Many of the experiments involve long-range fields which may in principle
alter the universality class -- perhaps in ways that differ from that
of the infinite--range model. Some experiments involve different kinds of
disorder, such as correlated disorder rather than point disorder and random
anisotropies or random bonds rather than random fields (which probably 
will not change the universality class). Also, there are new 
possible critical behaviors when one starts with other local order 
parameters: for example, if we start not with Ising spins but with
vectors~\cite{daSiKard99}.

\item {\bf Extracting exponents and scaling functions.}
We need a more organized approach, both experimentally and computationally,
for extracting critical exponents and reporting scaling functions. 
These hysteresis models represent a serious challenge, far more difficult
than most of the critical points studied in equilibrium phase transitions.
In part this is because the critical behavior is so much {\em more} important:
because the avalanches become large so far from $R_c$, one cannot get
close to $R_c$ without having severe finite size effects -- hence the
$R$ dependence of the parameters becomes a problem.

\end{enumerate}

\section*{Acknowledgments}

The work (and many of the discussions) reviewed in this article represent
collaborations with many of our students and colleagues. We acknowledge
support from NSF DMR-0218475, NSF DMR-0314279 and NSF DMR-9976550 
(Materials Computation Center). 

\appendix
\section{Derivation of the scaling forms and corrections}
\label{app:ScalingForms}

In our work we make extensive use of scaling collapses.%
  \footnote{This appendix follows closely the presentation 
  in~\cite{PerkDahmSeth96}.} 
Many variations are important to us: Widom scaling, finite-size scaling,
singular corrections to scaling, analytic corrections to scaling,
rotating axes, and exponentially diverging correlation length scaling.
The underlying theoretical framework for scaling is given by the
renormalization group,
in the context of equilibrium critical phenomena and by now well
explicated in a variety of texts.

We have discovered that we can derive all the scaling forms and
corrections that have been important to us from two simple hypotheses
(found in critical regions): universality and invariance under
reparameterizations. {\sl Universality} is the statement that two
completely different systems will behave the same near their critical
point (for example, they can have exactly the same kinds
of correlations). {\sl Reparameterization invariance} is the statement
that smooth changes in the units or methods of measurement should not
affect the critical properties -- or rather, that the predictions of 
our scaling theories are only valid up to a smooth change of coordinates
in the control parameters (the analytic corrections to scaling we discussed
in section~\ref{sec:Intro}). We use these properties to develop the
scaling forms and corrections we use in this paper. Each example we
cover will build on the previous ones while developing a new idea.

Our first example repeats our discussion of size versus duration
in section~\ref{sec:ExponentsScaling}.
Consider some property $F$ of a system at its
critical point, as a function of a scale $x$.  $F$ might be the
spin-spin correlation function as a function of distance $x$ (or it
might be the avalanche probability distribution as a function of size
$x$, etc.)  If two different experimental systems are at the same
critical point, their $F$'s must agree.  It would seem clear that they
cannot be expected to be equal to one another: the overall scale of $F$
and the scale of $x$ will depend on the microscopic structure of the
materials.  The best one could imagine would be that
\begin{equation}
F_1(x_1) = A F_2( B x_2)
\label{eq:single}
\end{equation}
where $A$ would give the ratio of, say, the squared magnetic moment per
domain of the two materials, and $B$ gives the ratio of the domain sizes.

Now, consider comparing a system with itself, but with a different
measuring apparatus.  Universality in this self-referential sense would
imply $F(x) = A F(B x)$, for suitable $A$ and $B$.  If instead of using
finite constant $A$ and $B$, we arrange for an infinitesimal change in
the measurement of length scales, we find:
\begin{equation}
F(x) = (1-\alpha \epsilon)\ F\Bigl((1-\epsilon) x\Bigr)
\label{eq:small_single}
\end{equation}
where $\epsilon$ is small and $\alpha$ is some constant. Taking the
derivative of both sides with respect to $\epsilon$ and evaluating it at
$\epsilon = 0$, we find $-\alpha F = x F'$, so
\begin{equation}
F(x) \sim x^{-\alpha}.
\label{eq:power}
\end{equation}
The function $F$ is a power--law! The underlying reason why power--laws
are seen at critical points is that power laws look the same at
different scales.

Now consider a new measurement with a distorted measuring apparatus. Now
$F(x) \sim {\cal A}\Bigl[F\Bigl({\cal B}(x)\Bigr)\Bigr]$ where ${\cal
A}$ and ${\cal B}$ are some nonlinear functions.  For example, one might
measure the number of microscopic domains $x$ flipped in an avalanche,
or one might measure the total acoustic power ${\cal B}(x)$ emitted
during the avalanche; these two ``sizes'' should roughly scale with one
another, but nonlinear amplifications will occur while the spatial
extent of the avalanche is small compared to the wavelength of sound
emitted: we expand ${\cal B}(x) = B x + b_0 + b_1/x + \ldots$ Similarly,
our microphone may be nonlinear at large sound amplitudes, or the
absorption of sound in the medium may be nonlinear: ${\cal A}(F) = A F +
a_2 F^2 + \ldots$ So,
\begin{eqnarray}
{\cal A}\Bigl[F\Bigl({\cal B}(x)\Bigr)\Bigr]\ \approx\ 
	~~~~~~~~~~~~~~~~~~~~~~~~~~~~~~~~~\nonumber \\
	  A \Bigl(F(B x) + F'(B x)(b_0 + b_1/x + \ldots)\ +\ \nonumber \\
		F''(B x)(\ldots) + \ldots \Bigr)
		+\ a_2 F^2(B x)\ + \ldots 
\label{eq:nonlinear_single}
\end{eqnarray}
We can certainly see that our assumption of universality cannot hold
everywhere: for large $F$ or small $x$ the assumption of
reparameterization invariance (\ref{eq:nonlinear_single}) prevents any
simple universal form.  Where is universality possible?  We can take the
power-law form $F(x) \sim x^{-\alpha} = x^{\log A/\log B}$ which is the
only form allowed by linear reparameterizations and plug it into
(\ref{eq:nonlinear_single}), and we see that all these nonlinear
corrections are subdominant ({\it i.e.}, small) for large $x$ and small
$F$ (presuming $\alpha>0$).  If $\alpha>1$, the leading correction is
due to $b_0$ and we expect $x^{-\alpha-1}$ corrections to the universal
power law at small distances; if $0<\alpha<1$ the dominant correction
is due to $a_2$, and we expect corrections of order $x^{-2\alpha}$.
Thus our assumptions of universality and reparameterization invariance
both lead us to the power-law scaling forms and inform us as to some
expected deviations from these forms.  Notice that the simple rescaling
led to the universal power-law predictions, and that the more complicated
nonlinear rescalings taught us about the dominant corrections: this will
keep happening with our other examples.

For our second example, let us consider a property $K$ of a system, as a
function of some external parameter $R$, as we vary $R$ through the
critical point $R_c$ for the material (so $r=R-R_c$ is small).  $K$
might represent the second moment of the avalanche size distribution,
where $R$ would represent the value of the randomness; alternatively $K$
might represent the fractional change in magnetization $\Delta M$ at the
infinite avalanche $\ldots$ If two different experimental systems are
both near their critical points ($r_1$ and $r_2$ both small), then
universality demands that the dependence of $K_1$ and $K_2$ on
``temperature'' $R$ must agree, up to overall changes in scale. Thus,
using a simple linear rescaling $K(r) = (1-\mu \epsilon)
K\Bigl((1-\epsilon) r\Bigr)$ leads as above to the prediction
\begin{equation}
K(r) = r^{-\mu}.
\label{eq:linear_single}
\end{equation}

Now let us consider nonlinear rescalings, somewhat different than the
one discussed above. In particular, the nonlinearity of our measurement
of $K$ can be dependent on $r$. So,
\begin{equation}
{\cal A}_r\Bigl(K(r)\Bigr) = a_0 + a_1 r + a_2 r^2 + \ldots +
			 a_{01} K(r) + \ldots
\label{eq:nonlinear_k}
\end{equation}
If $\mu>0$, these analytic corrections don't change the dominant power
law near $r=0$.  However, if $\mu<0$, all the terms $a_n$ for $n<-\mu$
will be more important than the singular term!  Only after fitting them
to the data and subtracting them will the residual singularity be
measurable. For the fractional change in magnetization: $\Delta M \sim
r^\beta$ has $0< \beta < 1$ (at least above three dimensions), so we
might think we need to subtract off a constant term $a_0$, but $\Delta M
= 0$ for $R \ge R_c$, so $a_0$ is zero. On the other hand, in a previous
paper\cite{DahmSeth96}, we discussed the singularity in the area of the
hysteresis loop: $Area \sim r^{2-\alpha}$, where $2-\alpha = \beta +
\beta \delta$ is an analogue to the specific heat in thermal systems.
Since $\alpha$ is near zero (slightly positive from our estimates of
$\beta$ and $\delta$ in 3, 4, and 5 dimensions), measuring it would
necessitate our fitting and subtracting three terms (constant, linear,
and quadratic in $r$): we did not measure the area for that reason.

For our third example, let's consider a function $F(x,r)$, depending on
both a scale $x$ and an external parameter $r$.  For example, $F$ might
be the probability $D_{int}$ that an avalanche of size $x$ will occur
during a hysteresis loop at disorder $r=R-R_c$.  Universality implies
that two different systems must have the same $F$ up to changes in
scale, and therefore that $F$ measured at one $r$ must have the same
form as if measured at a different $r$. To start with, we consider a
simple linear rescaling:
\begin{equation}
F(x,r) = (1 - \alpha \epsilon)\
	F\Bigl( (1-\epsilon) x, (1+\zeta \epsilon) r \Bigr).
\label{eq:linear_double}
\end{equation}
Taking the derivative of both sides with respect to $\epsilon$ gives a
partial differential equation that can be manipulated to show $F$ has a
scaling form.  Instead, we change variables to a new variable $y =
x^\zeta r$ (which satisfies $y'=y$ to order $\epsilon$).  If $\tilde
F(x,y) \equiv F(x,r)$ is our function measured in the new variables,
then
\begin{equation}
F(x,r) = \tilde F(x,y) = (1-\alpha \epsilon)\
		 \tilde F\Bigl((1-\epsilon)x,y\Bigr)
\label{eq:linear_double_tilde}
\end{equation}
and $-\alpha \tilde F = x \, \partial \tilde F/\partial x$ shows that at
fixed $y$, $F\sim x^{-\alpha}$, with a coefficient ${\cal F}(y)$ which
can depend on $y$.  Hence we get the scaling form
\begin{equation}
F(x,r) \sim x^{-\alpha}\ {\cal F}(x^\zeta r).
\label{eq:double_scaling}
\end{equation}
This is just Widom scaling. The critical exponents $\alpha$ and $\zeta$,
and the scaling function ${\cal F}(x^\zeta r)$ are universal (two
different systems near their critical point will have the {\sl same}
critical exponents and scaling functions). We don't need to discuss
corrections to scaling for this case, as they are similar to those
discussed above and below (and because none were dominant in our cases).

Notice that if we sit at the critical point $r=0$, the above result
reduces to equation (\ref{eq:power}) so long as ${\cal F}(0)$ is not zero
or infinity. If, on the other hand, ${\cal F}(y) \sim y^n$ as $y \to 0$,
the two-variable scaling function gives a singular correction to the
power--law near the critical point: $F(x,r) \sim x^{-\alpha}\ {\cal
F}(x^\zeta r) \sim x^{-\alpha + n \zeta}\ $ for $x <\!< r^{-1/\zeta}$:
only when $x \sim r^{-1/\zeta}$ will the power-law $x^{-\alpha}$ be
observed.  This is what happened in two dimensions to the integrated
avalanche size distribution and the avalanche correlation functions
(figures~35,~36, and 37b in reference~\cite{PerkDahmSeth96}).

%{\bf Discussion of "rotating axes" corrections to scaling for this case.
%Note exponential decay of $F$ makes for convergent integrals, subdominant
%corrections.  Correlation, int avalanche size dist, avalanche time dist,
%}

For the fourth example, we address finite-size scaling of a property $K$
of the system, as we vary a parameter $r$.  If we measure $K(r,L)$ for
a variety of sizes $L$ (say, all with periodic boundary conditions), we 
expect (in complete analogy to (\ref{eq:double_scaling}))
\begin{equation}
K(r,L) \sim r^{-\mu}\ {\cal K}(r L^{1/\nu}).
\label{eq:finite_size_scaling}
\end{equation}
Now, suppose our ``thermometer'' measuring $r$ is weakly size-dependent,
so the measured variable is ${\cal C}(r) = r + c/L + c_2/L^2 + \ldots$\ 
The effects on the scaling function is
\begin{eqnarray}
K\Bigl({\cal C}(r),L\Bigr) \sim r^{-\mu}\ \times \nonumber
	~~~~~~~~~~~~~~~~~~~~~ \\
	 \Bigl({\cal K}(r L^{1/\nu})\ +\ 
	~~~~~~~~~~~~~~ \nonumber \\ 
	 (c L^{1/\nu-1} + c_2 L^{1/\nu-2})\ {\cal K}'(r L^{1/\nu}) +
			\ldots \Bigr).
\label{eq:finite_size_corrections_to_scaling}
\end{eqnarray}

In two and three dimensions, $\nu>1$ and these correction terms are
subdominant. In four and five dimensions, we find $1/2 < \nu < 1$, so we
should include the term multiplied by $c$ in equation
(\ref{eq:finite_size_corrections_to_scaling}). However, we believe this
first term is zero for our problem. For a fixed boundary problem (all
spins ``up'' at the boundary) with a first order transition, there is
indeed a term like $c/L$ in $r(L)$. At a
critical transition, the leading correction to $r(L)$ can be $c/L$ or a
higher power of $L$ ($1/L^2$ and so on). This seems to depend on the
model studied, the geometry of the system, and the boundary conditions
(free, periodic, ferromagnetic, $\ldots$)
Furthermore, for the same kind of model, the coefficient $c$ itself
depends on the geometry and boundary conditions, and it can even vanish,
which leaves only higher order corrections. In a periodic boundary
conditions problem like ours, we expect that the correction is smaller
than $c/L$. Our finite-size scaling collapses for spanning avalanches
$N$, the second moments $\langle S^2\rangle$, and the magnetization jump
$\Delta M$, were successfully done by letting $c=0$.

For the fifth example, consider a property $K$ depending on two external
parameters: $r$ (the disorder for example) and $h$ (could be the
external magnetic field $H-H_c$). Analogous to (\ref{eq:double_scaling}),
$K$ should then scale as
\begin{equation} 
K(r,h) \sim r^{-\mu}\ {\cal K}(h/r^{\beta\delta}).
\label{eq:double_scaling_2}
\end{equation}
Consider now the likely dependence of the field $h$ on the disorder $r$.
A typical system will have a measured field which depends on the
randomness: $\tilde{\cal C}(h) = h +  b\, r + b_2 r^2 + \ldots$
(Corresponding nonlinearities in the effective value of $r$ are
subdominant.) This system will have
\begin{eqnarray}
\label{eq:scaling_2}
K\Bigl(r,\tilde{\cal C}(h)\Bigr)\ =\ r^{-\mu}\ \times 
	~~~~~~~~~~~~~~~~~~~~~~~~~~~~~~~~~~ \nonumber \\
	 \Bigl( {\cal K}(h/r^{\beta\delta}) 
	    + (b\, r + b_2 r^2)\ r^{-\beta\delta}\ 
		{\cal K}'(h/r^{\beta\delta}) \Bigr).
\label{eq:rotated_double_scaling}
\end{eqnarray}
Now, for our system $1 < \beta \delta < 2$ for dimensions three and
above. This means that the term multiplied by $b$ is dominant over the
critical scaling singularity: unless one shifts the measured $h$ to the
appropriate $h'=h + b\,r$, the curves will not collapse ({\it e.g.}, the
peaks will not line up horizontally).  We measure this (non-universal)
constant for our system using the derivative of
the magnetization with field $dM/dH(r,h)$.  The magnetization $M(r,h)$
and the correlation length $\xi(r,h)$ should also collapse according to
equation (\ref{eq:double_scaling_2}) (but with $h + b\,r$ instead of $h$);
we don't directly measure the correlation length, and the collapse of
$M(r,h)$ in the bottom figure~\ref{fig:MofH} includes the effects of the tilt
$b$.  In two dimensions, $\beta \delta$ is large (probably infinite), so
in principle we should need an infinite number of correction terms: in
practice, we tried lining up the peaks in the curves (with no correction
terms); because we did not know $\beta$ (which we usually obtained
from $\Delta M$, which gives $\beta/\nu=0$ in two dimensions), we failed
to extract reliable exponents in two dimensions from $dM/dH$.

For the sixth example, suppose $F$ depends on $r$, $h$, and a size $x$.
Then from the previous analysis, we expect
\begin{equation}
F(x,r,h) \sim x^{-\alpha}\ {\cal F}(x^\zeta r,\, h/r^{\beta \delta}).
\label{eq:triple_scaling}
\end{equation}
Notice that universality only removes one variable from the scaling
form. One could in practice do two--variable scaling collapses (and we
believe someone has probably done it), but for our purposes these more
general scaling forms are used by fixing one of the variables.  For
example, we measure the avalanche size distribution at various values of
$h$ (binned in small ranges), at the critical disorder $r=0$.  We can
make sense of equation (\ref{eq:triple_scaling}) by changing variables from
$h/r^{\beta \delta}$ to $x^{\zeta \beta \delta} h$:
\begin{equation}
F(x,r,h) \sim x^{-\alpha} \tilde{\cal F}(x^\zeta r,\, x^{\zeta \beta \delta} h).
\label{eq:triple_scaling_nice}
\end{equation}
Before we can set $r=0$, we must see what are the possible corrections
to scaling in this case. If the disorder $r$ depends on the field, then
instead of the variable $r$, we must use $r + a h$ (the analysis is
analogous to the one in example five; other corrections are
subdominant). Setting $r=0$ now, leaves $F$ dependent on its first
variable, as well as the second:
\begin{eqnarray}
F(x,r,h) & \sim\  x^{-\alpha}\ \tilde{\cal F}(x^\zeta (a h),\,
			x^{\zeta \beta \delta} h)
          \approx\  x^{-\alpha}\ \times \nonumber \\
	 &  \Bigl(\tilde{\cal F}(0,\ x^{\zeta \beta \delta} h) + 
	~~~~~~~~~~~~~~~~~~~~~ \nonumber \\  
	&  a h x^\zeta\
	 \tilde{\cal F}^{(1,0)}(0,\, x^{\zeta \beta \delta} h) \Bigr),
\label{eq:triple_scaling_reduced_corrections}
\end{eqnarray}
where $\tilde{\cal F}^{(1,0)}$ is the derivative of $\tilde{\cal F}$
with respect to the first variable (keeping the second fixed).

For the binned avalanche size distribution, $x^\zeta$ is $S^\sigma$,
where $0 \le \sigma < 1/2$ as we move from two dimensions to five and
above. Thus, the correction term will only be important for rather large
avalanches, $S > h^{-1/\sigma}$, so long as we are close to the critical
point.  Expressed in terms of the scaling variable, important
corrections to scaling occur if the scaling variable $X =
S^{\sigma\beta\delta} h >  h^{1-\beta\delta}$. For us, $\beta \delta >
3/2$, and we only use fields near the critical field ($h < 0.08$), so
the corrections will become of order one when $X=4$ for the largest $h$
we use. In $3$ and $4$ dimensions, this correction does not affect our
scaling collapses, while in $5$ dimensions some of the data needs this
correction. We have tried to avoid this problem (since we don't measure
our data such that it can be used in a two--variable scaling collapse)
by concentrating on collapsing the regions in our data curves where this
correction is negligible.

A similar analysis can be done for the avalanche time distribution,
which has two ``sizes'' $S$ and $t$ and one parameter $r$ which is set
to zero; because we integrate over the field $h$ the correction in
(\ref{eq:triple_scaling_reduced_corrections}) does not occur, and other
scaling corrections are small.

Finally, we discuss the unusual exponential scaling forms we developed
to collapse our data in two dimensions. If we assume that the critical
disorder $R_c$ is zero {\it and} that the linear term in the rescaling
of $r$ vanishes ($\zeta \epsilon r$ in equation\ (\ref{eq:linear_double})
vanishes), then from symmetry the correction has to be cubic, and
equation\ (\ref{eq:linear_double}) becomes:
\begin{equation}
F(x,r) = (1 - \alpha \epsilon)\
        F\Bigl( (1-\epsilon) x,\, (1+ k \epsilon\, r^2) r \Bigr).
\label{eq:exponential_scaling_cubic}
\end{equation}
with $k$ (which is not universal) and $\alpha$ constants, and $\epsilon$
small.

Taking the derivative of both sides with respect to $\epsilon$ and
setting it equal to zero gives a partial differential equation for the
function $F$. To solve for $F$, we do a change of variable: $(x,r)
\rightarrow (x,y)$ with $y=x\ e^{-a^*/r^2}$. The constant $a^*$ is
determined by requiring that $y$ rescales onto itself to order
$\epsilon$: we find $a^*=1/2\,k$. We then have:
\begin{equation}
0 = -\alpha\ \tilde{F}(x,y) - {\partial{\tilde{F}} \over \partial{x}}\ x
\label{eq:exponential_scaling_cubic_2}
\end{equation}
which gives
\begin{equation}
F(x,r) = x^{-\alpha}\ \tilde{\cal{F}}\Bigl(xe^{-1/2\,k\,r^2}\Bigr).
\label{eq:exponential_scaling_cubic_3}
\end{equation}

This is one of the forms we use in $2$ dimensions for the scaling
collapse of the second moments $\langle S^2 \rangle_{int}$, the
avalanche size distribution $D_{int}$ integrated over the field $H$, the
avalanche correlation $G_{int}$, and the spanning avalanches $N$. We use
another form too which is obtained by assuming that the critical
disorder $R_c$ is not zero but that the linear term in the rescaling of
$r$ still vanishes. Instead of equation\
(\ref{eq:exponential_scaling_cubic}), we have:
\begin{equation}
F(x,r) = (1 - \alpha \epsilon)\
        F\Bigl( (1-\epsilon) x,\, (1+ \ell \epsilon\, r) r \Bigr).
\label{eq:exponential_scaling_square}
\end{equation}
The function $F$ becomes:
\begin{equation}
F(x,r) = x^{-\alpha}\ \tilde{\cal{F}}\Bigl(xe^{-1/\ell\,r}\Bigr).
\label{eq:exponential_scaling_square_2}
\end{equation}
The corrections to scaling for the last two forms (equations\
(\ref{eq:exponential_scaling_cubic_3}) and
(\ref{eq:exponential_scaling_square_2})) are similar to the ones discussed
above. They are all are subdominant.

\bibliographystyle{hystbk}
\bibliography{RFIM}

\end{document}